\documentclass[aps,prl,reprint,twocolumn,superscriptaddress,floatfix,nofootinbib,longbibliography]{revtex4-2}
\usepackage{graphicx,amsmath,amsfonts,amssymb,amsthm,xr}
\usepackage{epsfig,amsmath,amssymb,color,dsfont,upgreek,physics}
\usepackage{mathrsfs}
\usepackage{mathtools}
\usepackage{adjustbox}
\usepackage{bbold}
\usepackage{float}
\usepackage[caption=false]{subfig}
\usepackage{multirow}

\usepackage[bookmarks=true,colorlinks,linkcolor=OrangeRed,urlcolor=NavyBlue,citecolor=RoyalBlue]{hyperref}
\usepackage[dvipsnames]{xcolor}
\usepackage{orcidlink}

\usepackage{graphicx}
\usepackage{dcolumn}
\usepackage{bm}
\usepackage{color}

\definecolor{mypurple}{rgb}{0.49,0.18,0.56}

\usetikzlibrary{arrows.meta}
\usepackage{changes}
\usetikzlibrary{quantikz2}
\usepackage{helvet}
\usepackage[tracking=true]{microtype}
\SetTracking{encoding=*}{90} % 100 = 10% spacing, adjust as you like

\begin{document}
\title{Observation of genuine $2+1$D string dynamics in a U$(1)$ lattice gauge theory with a tunable plaquette term on a trapped-ion quantum computer}

\author{Rohan Joshi${}^{\orcidlink{0000-0003-1520-7146}}$}
\thanks{These authors contributed equally to this work.}
\affiliation{Max Planck Institute of Quantum Optics, 85748 Garching, Germany}
\affiliation{Department of Physics and Arnold Sommerfeld Center for Theoretical Physics (ASC), Ludwig Maximilian University of Munich, 80333 Munich, Germany}
\affiliation{Munich Center for Quantum Science and Technology (MCQST), 80799 Munich, Germany}

\author{Yizhuo Tian${}^{\orcidlink{0009-0006-7012-3358}}$}
\thanks{These authors contributed equally to this work.}
\affiliation{Department of Physics and Arnold Sommerfeld Center for Theoretical Physics (ASC), Ludwig Maximilian University of Munich, 80333 Munich, Germany}
\affiliation{Munich Center for Quantum Science and Technology (MCQST), 80799 Munich, Germany}

\author{Kevin Hemery${}^{\orcidlink{0000-0001-7086-593X}}$}
\affiliation{Quantinuum, Leopoldstr.~180, 80804 Munich, Germany}

\author{Kaidi Xu${}^{\orcidlink{0000-0003-2184-0829}}$}
\affiliation{Max Planck Institute of Quantum Optics, 85748 Garching, Germany}
\affiliation{Department of Physics and Arnold Sommerfeld Center for Theoretical Physics (ASC), Ludwig Maximilian University of Munich, 80333 Munich, Germany}
\affiliation{Munich Center for Quantum Science and Technology (MCQST), 80799 Munich, Germany}

\author{N.~S.~Srivatsa${}^{\orcidlink{0000-0001-6433-450X}}$}
\affiliation{Max Planck Institute of Quantum Optics, 85748 Garching, Germany}
\affiliation{Department of Physics and Arnold Sommerfeld Center for Theoretical Physics (ASC), Ludwig Maximilian University of Munich, 80333 Munich, Germany}
\affiliation{Munich Center for Quantum Science and Technology (MCQST), 80799 Munich, Germany}

\author{Jesse J.~Osborne${}^{\orcidlink{0000-0003-0415-0690}}$}
\affiliation{Max Planck Institute of Quantum Optics, 85748 Garching, Germany}
\affiliation{Department of Physics and Arnold Sommerfeld Center for Theoretical Physics (ASC), Ludwig Maximilian University of Munich, 80333 Munich, Germany}
\affiliation{Munich Center for Quantum Science and Technology (MCQST), 80799 Munich, Germany}

\author{Henrik Dreyer${}^{\orcidlink{0000-0002-1480-6406}}$}
\affiliation{Quantinuum, Leopoldstr.~180, 80804 Munich, Germany}

\author{Enrico Rinaldi${}^{\orcidlink{0000-0003-4134-809X}}$}
\affiliation{Quantinuum, Partnership House, Carlisle Place, London SW1P 1BX, UK}
\affiliation{RIKEN Center for Quantum Computing (RQC), RIKEN, Wako, Saitama 351-0198, Japan}
\affiliation{School of Mathematical Sciences, Queen Mary University of London, Mile End Road, London, E1 4NS, UK}

\author{Jad C.~Halimeh${}^{\orcidlink{0000-0002-0659-7990}}$}
\email{jad.halimeh@lmu.de}
\affiliation{Department of Physics and Arnold Sommerfeld Center for Theoretical Physics (ASC), Ludwig Maximilian University of Munich, 80333 Munich, Germany}
\affiliation{Max Planck Institute of Quantum Optics, 85748 Garching, Germany}
\affiliation{Munich Center for Quantum Science and Technology (MCQST), 80799 Munich, Germany}
\affiliation{Department of Physics, College of Science, Kyung Hee University, Seoul 02447, Republic of Korea}

\begin{abstract}
Quantum simulations of high-energy physics in $2+1$D can probe dynamical phenomena nonexistent in one spatial dimension and access regimes that are challenging for existing classical simulation methods. 
For string dynamics---relevant to hadronization---a plaquette term is required to realize genuine $2+1$D behavior, as it endows the gauge field with dynamics and enables the propagation of photon-like excitations. 
Here, we realize a U$(1)$ quantum link model of quantum electrodynamics in two spatial dimensions with a tunable plaquette term on a \texttt{Quantinuum System Model H2} quantum computer. 
We implement, to our knowledge, the largest quantum simulation of string-breaking dynamics reported to date, on a \(5 \times 4\) matter-site square lattice using \(51\) qubits. 
The simulation uses a shallow circuit design with a two-qubit gate depth of $28$ per Trotter step and up to $1540$ entangling gates. 
Starting from far-from-equilibrium string configurations, we measure the probability for the string to propagate within the lattice plane and find signatures of genuine $2+1$D dynamics only when the plaquette term is present. In a resonant regime, we observe the annihilation of string segments accompanied by the production of electron--positron pairs that screen them. 
We further find that, only with a nonzero plaquette term, matter creation extends across the lattice plane rather than remaining confined to the initial string path. 
These results experimentally realize string breaking and demonstrate the emergence of dynamical gauge fields in two spatial dimensions, establishing a route to photon-like propagation in programmable quantum simulators of gauge theories.
\end{abstract}

\date{\today} 
\maketitle

Gauge theories provide the fundamental framework for describing elementary interactions in nature, from electromagnetism to the strong force \cite{Weinberg1995QuantumTheoryFields,Weinberg:2004kv,peskin2018introduction}. They constitute the Standard Model of particle physics, encapsulating, e.g., quantum electrodynamics (QED) and quantum chromodynamics, and are governed by local gauge symmetries, which give rise to rich collective phenomena including confinement, flux-tube formation, and nonperturbative dynamics of gauge fields \cite{Ellis2003QCDColliderPhysics}. Understanding such phenomena remains a major challenge because the real-time evolution of strongly interacting gauge theories is difficult to access with classical computational methods.

Lattice gauge theories (LGTs) offer a robust and systematic framework to study gauge theories from first principles by discretizing spacetime and representing gauge fields on the links of a lattice \cite{Kogut1975HamiltonianFormulationWilsons,Kogut1979AnIntroductionToLatticeGaugeTheory,Rothe2012LatticeGaugeTheories}. Originally invented to study quark confinement \cite{Wilson1974ConfinementQuarks,Wilson1977QuarksStringsLattice}, LGTs have also become extremely successful in condensed matter physics as effective descriptions of exotic phases such as quantum spin liquids \cite{wen2004quantum,Balents2010SpinLiquidsFrustrated,Savary2016QuantumSpinLiquids} and the fractional quantum Hall effect \cite{Kleinert1989GaugeFieldsCondensed,Fradkin2013FieldTheoriesCondensed}, as well as in quantum many-body physics, where they provide a venue of rich nonergodic dynamics~\cite{Smith2017DisorderFreeLocalization,Brenes2018ManyBodyLocalization,Smith2017AbsenceOfErgodicity,Karpov2021DisorderFreeLocalization,Sous2021PhononInducedDisorder,Chakraborty2022DisorderFreeLocalization,Halimeh2022EnhancingDisorderFreeLocalization,Surace2020LatticeGaugeTheories,Lang2022DisorderFreeLocalization,Desaules2023WeakErgodicityBreaking,Desaules2023ProminentQuantumManyBodyScars,Aramthottil2022ScarStates,Tarabunga2023ManyBodyMagic,Desaules2024ergodicitybreaking,Desaules2024MassAssistedLocalDeconfinement,Hudomal2022DrivingQuantumManyBodyScars,Jeyaretnam2025HilbertSpaceFragmentation,Smith2025Nonstabilizerness,Falcao2025nonstabilizerness,Esposito2025magicdiscretelatticegaugetheories,Ciavarella2025GenericHilbertSpaceFragmentation,Ciavarella:2025tdl,Steinegger2025GeometricFragmentationAnomalousThermalization,Ebner2024EntanglementEntropy,Halimeh2023robustquantummany,Iadecola2020QuantumManyBodyScar,Banerjee2021QuantumScarsZeroModes,Biswas2022ScarsFromProtectedZeroModes,Daniel2023BridgingQuantumCriticality,Sau2024sublatticescarsbeyond,Osborne2024QuantumManyBodyScarring,Budde2024QuantumManyBodyScars,Calajo2025QuantumManyBodyScarringNonAbelian,Hartse2025StabilizerScars,cataldi2025disorderfreelocalizationfragmentationnonabelian}. The Hamiltonians of these models typically contain three distinct types of contributions: electric-field terms associated with the links, magnetic terms that couple the gauge fields around elementary plaquettes of the lattice, and a minimal coupling term for gauge-invariant matter dynamics. While the electric terms govern the energy stored in flux lines, the magnetic plaquette term enables the fluctuation and motion of gauge flux, and is therefore the key ingredient that generates genuinely higher-dimensional gauge dynamics \cite{Rothe2012LatticeGaugeTheories}.

\begin{figure*}
\includegraphics[width=\linewidth]{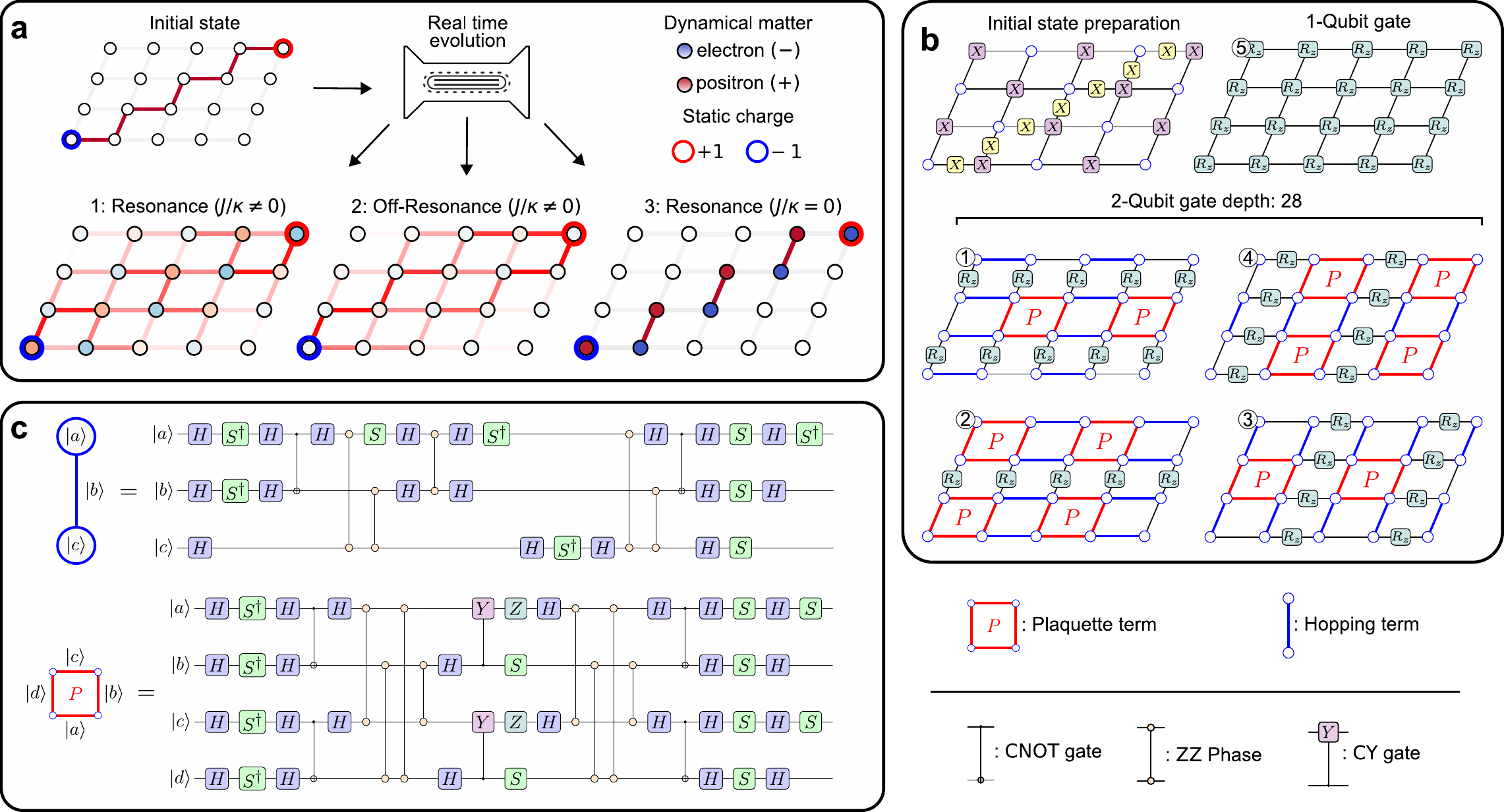}
\caption{\textbf{Quantum simulation of genuine $2+1$D string dynamics in a U$(1)$ lattice gauge theory on a $5 \times 4$ square lattice with open boundary conditions.} $\textbf{\textsf{a}}$  Illustrations of the three dynamical regimes probed after starting from an initial diagonal string configuration time-evolved by \texttt{Quantinuum System Model H2}: 1. Resonant and 2. off-resonant dynamics in the presence of a plaquette term (both hardware data), and 3. resonant dynamics in the absence of a plaquette term (TN simulation here, hardware data below). $\textbf{\textsf{b}}$ Implementation of the LGT Hamiltonian~\eqref{eq:H_QLM} on a quantum circuit. The initial diagonal string state is prepared through the application of $X$ gates on the relevant matter qubits and links. Each Trotter step is decomposed into four sublayers composed of mutually commuting hopping and plaquette terms, excluding the mass term, followed by a single-qubit rotation layer implementing the mass term. $\textbf{\textsf{c}}$  Circuit decompositions of the hopping and plaquette terms.}
\label{fig:Schematic}
\end{figure*}

In the absence of such plaquette interactions, the dynamics of gauge fields on higher-dimensional lattices can be effectively described by $1+1$D processes, even though the underlying geometry has more than one spatial dimension \cite{Tian2025RolePlaquetteTerm}. As a result, experimentally realizing tunable plaquette interactions constitutes a crucial step toward exploring the full dynamical richness of gauge theories in more than one spatial dimension \cite{Zohar2021QuantumSimulationLattice}. Achieving this goal would enable controlled investigations of phenomena relevant to collider physics such as flux fluctuations, string dynamics, and the collective behavior of gauge fields beyond effectively $1+1$D geometries \cite{Halimeh2025QuantumSimulationOutofequilibrium}.

Classical techniques face serious difficulties in probing out-of-equilibrium LGT dynamics. Monte Carlo techniques, which have enabled deep insights into LGT physics in equilibrium \cite{Creutz1979MonteCarloStudy,Creutz1980MonteCarloStudy,Creutz1983MonteCarloComputations,Creutz1988LatticeGaugeTheory,Creutz1989LatticeGaugeTheories,montvay1994quantum,Kieu1994MonteCarloSimulations,Hackett2019DigitizingGaugeFields}, encounter the sign problem in out-of-equilibrium settings \cite{Trottier1999StringBreakingDynamical,Nagata2022FinitedensityLatticeQCD}. Tensor network (TN) methods \cite{Schollwoeck2005DensityMatrixRenormalizationGroup,Schollwock2011DensitymatrixRenormalizationGroup,Orus2014PracticalIntroductionTensorNetworks,Orus2019TensorNetworksComplex,Paeckel2019TimeevolutionMethodsMatrixproduct,Montangero2018IntroductionTensorNetwork} are excellent for probing relatively long evolution times in one spatial dimension, but due to rapid growth of quantum entanglement, they are considerably challenged in higher spatial dimensions \cite{Magnifico2024TensorNetworksLattice}. 

\begin{figure*}
    \centering
    \includegraphics[width=\linewidth]{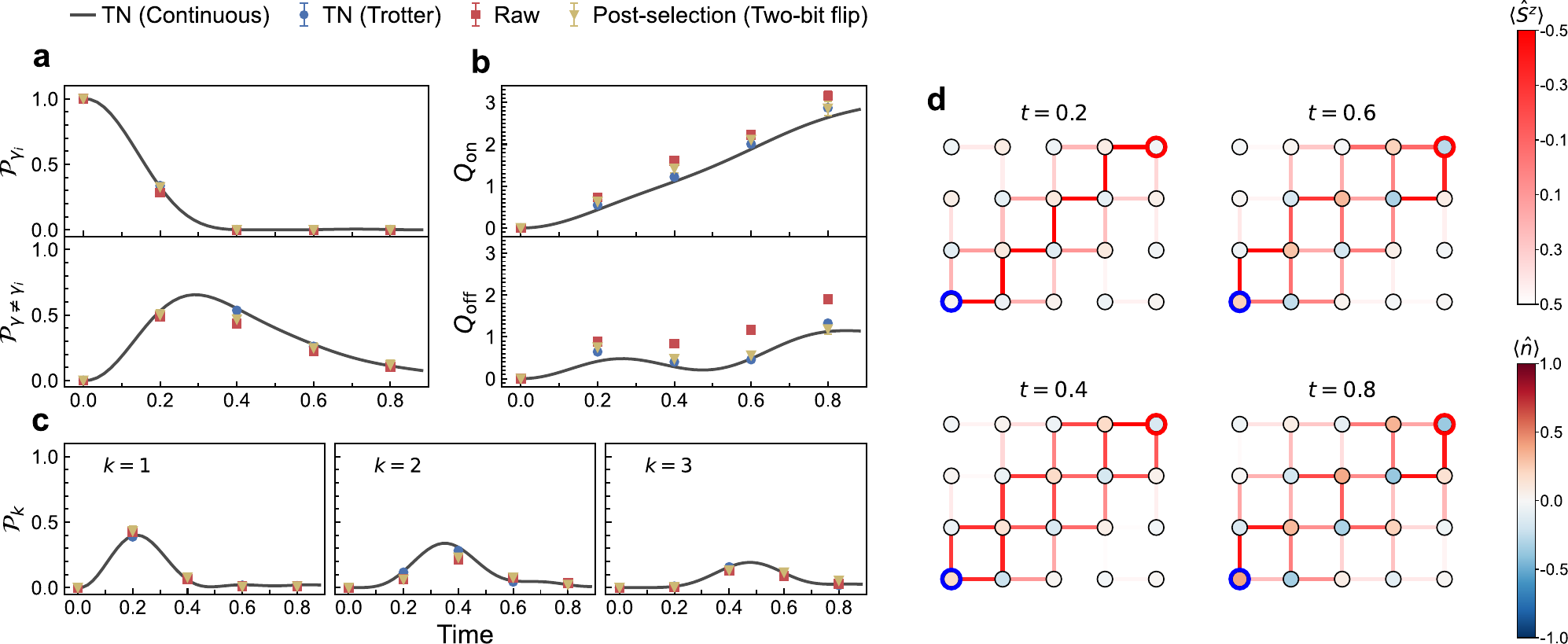}
    \caption{\textbf{Genuine $2+1$D string-breaking dynamics with a plaquette term.}
    On-resonant quench of a diagonal string on a $5 \times 4$ lattice with $\kappa = 1$, $m = 3$, $g = 6$, and $J=2$. \textbf{\textsf{a}} Probability of the initial string configuration, $\mathcal{P}_{\gamma_\mathrm{i}}$, and of all other string configurations combined, $\mathcal{P}_{\gamma\neq\gamma_\mathrm{i}}$, as a function of time. We compare the raw and two-bit flip post-selected hardware data with the continuous TN simulation and the TN simulation of the Trotterized circuit using the \texttt{Qiskit} MPS simulator. \textbf{\textsf{b}} Total number of dynamically proliferated charges on the initial string, $Q_{\mathrm{on}}$, and off the initial string, $Q_{\mathrm{off}}$. \textbf{\textsf{c}} Detailed distribution of the different string configurations generated by the local plaquette term acting at first, second, and third order. \textbf{\textsf{d}} Spatial distribution of the dynamical charge excitation $\langle \hat{n} \rangle$ and the link observable $\langle \hat{S}^z\rangle$ at four different Trotter steps obtained for the two-bit flip post-selection data.}
    \label{fig:on_resonance_plaq}
\end{figure*}

Quantum simulators provide a promising route to realize such LGT dynamics experimentally \cite{Byrnes2006SimulatingLatticeGauge, Dalmonte2016LatticeGaugeTheory, Zohar2015QuantumSimulationsLattice, Aidelsburger:2021mia, Zohar2021QuantumSimulationLattice, 
Barata2022MediumInducedJetBroadening,Klco2022StandardModelPhysics,Barata2023QuantumSimulationInMediumQCDJets,Barata2023RealTimeDynamicsofHyperonSpin, Bauer2023QuantumSimulationHighEnergy, Bauer2023QuantumSimulationFundamental,
DiMeglio2024QuantumComputingHighEnergy, Cheng2024EmergentGaugeTheory, Halimeh2022StabilizingGaugeTheories, Cohen2021QuantumAlgorithmsTransport,Barata2025ProbingCelestialEnergy, Lee2025QuantumComputingEnergy, Turro2024ClassicalQuantumComputing,Halimeh2023ColdatomQuantumSimulators,Bauer2025EfficientUseQuantum,Halimeh2025QuantumSimulationOutofequilibrium}. By directly incorporating the wave function, they naturally handle the growth of quantum entanglement \cite{Bauer2023QuantumSimulationFundamental,DiMeglio2024QuantumComputingHighEnergy}, bypassing the challenges faced by classical methods. Recent advances have enabled the implementation of LGTs on programmable quantum platforms, opening the possibility of probing nonperturbative phenomena in real time \cite{Martinez2016RealtimeDynamicsLattice, Klco2018QuantumclassicalComputationSchwinger,Gorg2019RealizationDensitydependentPeierls, Schweizer2019FloquetApproachZ2, Mil2020ScalableRealizationLocal, Yang2020ObservationGaugeInvariance, Wang2022ObservationEmergent$mathbbZ_2$, Su2023ObservationManybodyScarring, Zhou2022ThermalizationDynamicsGauge, Wang2023InterrelatedThermalizationQuantum, Zhang2025ObservationMicroscopicConfinement, Zhu2024ProbingFalseVacuum, Ciavarella2021TrailheadQuantumSimulation, Ciavarella2022PreparationSU3Lattice, Ciavarella2023QuantumSimulationLattice-1, Ciavarella2024QuantumSimulationSU3, 
Gustafson2024PrimitiveQuantumGates, Gustafson2024PrimitiveQuantumGates-1, Lamm2024BlockEncodingsDiscrete, Farrell2023PreparationsQuantumSimulations-1, Farrell2023PreparationsQuantumSimulations, 
Farrell2024ScalableCircuitsPreparing,
Farrell2024QuantumSimulationsHadron, Li2024SequencyHierarchyTruncation, Zemlevskiy2025ScalableQuantumSimulations, Lewis2019QubitModelU1, Atas2021SU2HadronsQuantum, ARahman:2022tkr, Atas2023SimulatingOnedimensionalQuantum, Mendicelli2023RealTimeEvolution, Kavaki2024SquarePlaquettesTriamond, Than2024PhaseDiagramQuantum, Angelides:2023noe, Gyawali2025ObservationDisorderfreeLocalization,  
Mildenberger2025Confinement$$mathbbZ_2$$Lattice, Schuhmacher2025ObservationHadronScattering, Davoudi2025QuantumComputationHadron, Saner2025RealTimeObservationAharonovBohm, Xiang2025RealtimeScatteringFreezeout, Wang2025ObservationInelasticMeson,li2025frameworkquantumsimulationsenergyloss,mark2025observationballisticplasmamemory,froland2025simulatingfullygaugefixedsu2,Hudomal2025ErgodicityBreakingMeetsCriticality,hayata2026onsetthermalizationqdeformedsu2,Cochran2025VisualizingDynamicsCharges, Gonzalez-Cuadra2025ObservationStringBreaking, Crippa2024AnalysisConfinementString, De2024ObservationStringbreakingDynamics, Liu2024StringBreakingMechanism, Alexandrou:2025vaj,Cobos2025RealTimeDynamics2+1D,ilcic2026observationrobustcoherentnonabelian,chen2026thermalizationsu2latticegauge, Balaji:2025yua, Balaji:2025afl}. However, the controlled realization of an LGT with a tunable magnetic plaquette interaction---essential for genuine $2+1$D LGT dynamics---has remained a central experimental challenge. Indeed, analog quantum simulation platforms struggle with implementing a plaquette term as it requires multi-body interactions that are challenging to engineer \cite{Halimeh2023ColdatomQuantumSimulators,Dai2017FourBodyRingExchange,Homeier2023RealisticScheme,Paredes2008minimuminstances}. This has prevented analog quantum simulation implementations from realizing plaquette terms in recent realizations of $2+1$D LGTs \cite{Gonzalez-Cuadra2025ObservationStringBreaking}, rendering their string dynamics effectively $1+1$D \cite{Tian2025RolePlaquetteTerm}. Digital platforms also face challenges due to limited gate depths and noise, making most implementations in $2+1$D without plaquette terms \cite{Gyawali2025ObservationDisorderfreeLocalization,Cobos2025RealTimeDynamics2+1D} with a few exceptions \cite{Cochran2025VisualizingDynamicsCharges}.

In this work, we realize a $2+1$D U$(1)$ LGT with a tunable plaquette term on a trapped-ion quantum computer; see Fig.~\ref{fig:Schematic}. We achieve this by implementing time evolution using a first-order Trotter decomposition with a structured circuit design over the lattice that, instead of applying Hamiltonian terms sequentially, groups commuting terms into parallel sublayers. This parallelization significantly reduces the overall circuit depth, as multiple commuting Hamiltonian terms can be implemented simultaneously within each sublayer. Moreover, this structured decomposition ensures that the two-qubit gate depth per Trotter step remains constant, independent of system size. Consequently, each Trotter step attains a low circuit depth, rendering the approach well suited for near-term quantum devices. After preparing initial product states comprised of far-from-equilibrium strings between two static charges, we quench the system and observe rich genuinely $2+1$D string dynamics. We tune a resonance between particle mass and a confining potential to instigate string breaking. This manifests in a segment of the string becoming screened by the formation of an electron--positron pair at the matter sites at its ends, thereby \textit{breaking} the string along that segment. Away from resonance, we show robust string fluctuations making full use of the two spatial dimensions. Our implementation hosts explicit matter and gauge degrees of freedom on a $5\times4$ square lattice, making it the largest known physical realization in $2+1$D to date, and the first scalable one with a tunable plaquette term.

\begin{figure*}
    \centering
    \includegraphics[width=\linewidth]{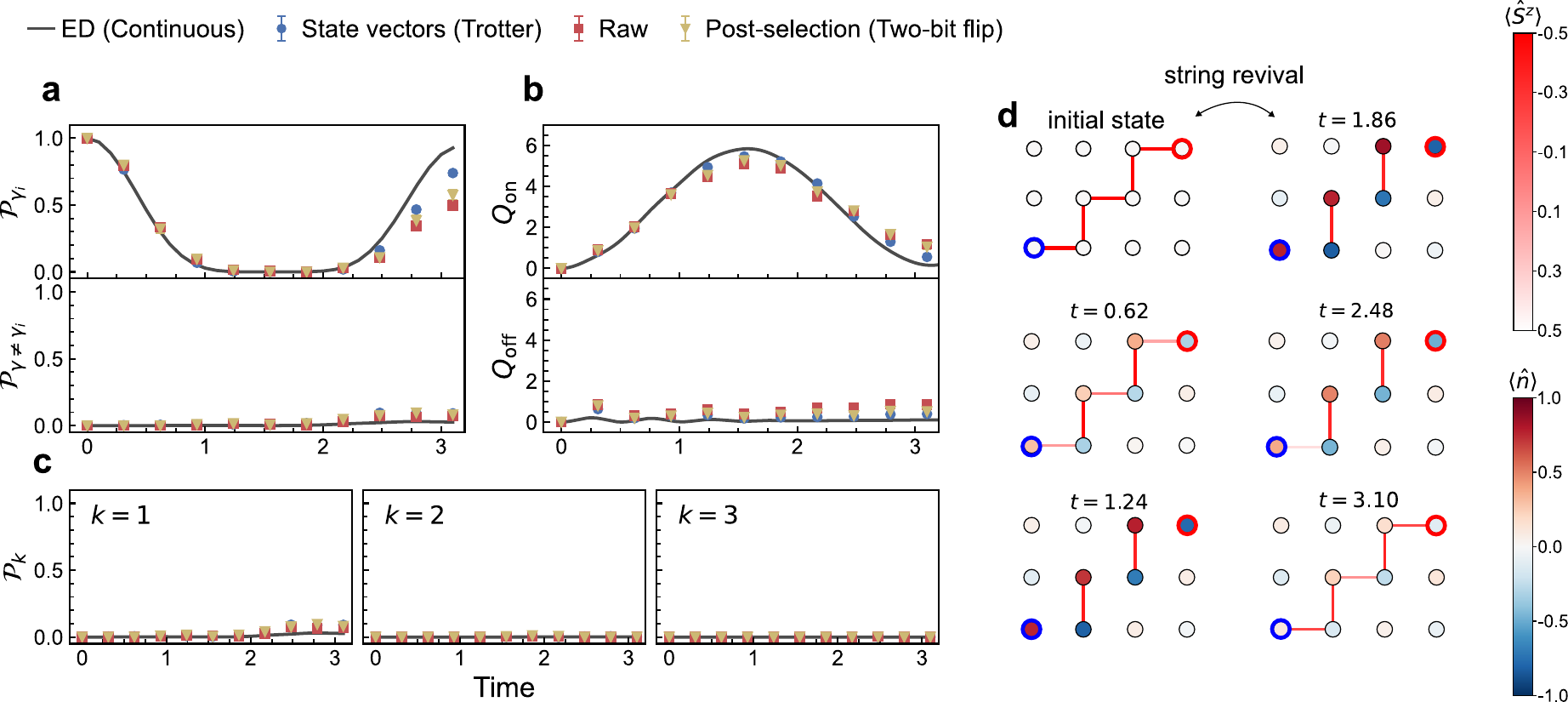}
    \caption{\textbf{Effective $1+1$D string dynamics in the absence of a plaquette term}.
    On-resonant quench of a diagonal string on a $4 \times 3$ lattice in the absence of the plaquette term ($J = 0$), with parameters $\kappa = 1$, $m = 3$, and $g = 6$. \textbf{\textsf{a}} Probability of the initial string configuration, $\mathcal{P}_{\gamma_\mathrm{i}}$, and of all other string configurations combined, $\mathcal{P}_{\gamma\neq\gamma_\mathrm{i}}$. We compare both raw and two-bit flip post-selected hardware data against continuous-time exact diagonalization (ED) simulations and \texttt{Qiskit} state-vector simulations of the Trotterized circuit. \textbf{\textsf{b}} Total number of dynamically proliferated charges on the initial string, $Q_{\mathrm{on}}$, and off the initial string, $Q_{\mathrm{off}}$. \textbf{\textsf{c}} Detailed distribution of the different string configurations generated by the local plaquette term acting at first, second, and third order. \textbf{\textsf{d}} Spatial distribution of the dynamical charge excitation $\langle \hat{n} \rangle$ and the link observable $\langle \hat{S}^z\rangle$ at different times obtained from the two-bit flip post-selection result.
    }
    \label{fig:on_resonance_no_plaq}
\end{figure*}

\textbf{Model.---}
We consider a $2+1$D U$(1)$ LGT on a square lattice with hardcore bosonic matter, a discretization of scalar QED, using the quantum link model formulation to truncate the gauge fields~\cite{Chandrasekharan1997QuantumLinkModels,Wiese2013UltracoldQuantumGases}. The Hamiltonian describing this theory is 

\begin{align}\nonumber
    \hat{H} =&\underbrace{-\kappa\sum_{\mathbf{j},\mu}\Big(s_{\mathbf{j},\mathbf{e}_{\mu}}
\hat{\phi}^{\dagger}_{\mathbf{j}}
\hat{S}^{+}_{\mathbf{j},\mathbf{e}_{\mu}}
\hat{\phi}_{\mathbf{j}+\mathbf{e}_{\mu}}
+\textrm{H.c.}\Big)}_{\hat{H}_\kappa} + \underbrace{m\sum_\mathbf{j} s_{\mathbf{j}}\hat{\phi}_\mathbf{j}^{\dag}\hat{\phi}_\mathbf{j}}_{\hat{H}_m}\\
    & - \underbrace{g\sum_{\mathbf{j}, \mathbf{\mu}}\hat{S}_\mathbf{j, e_{\mu}}^z}_{\hat{H}_E} + \underbrace{J \sum_{\square}\Big(\hat{U}_{\square}+\hat{U}_{\square}^{\dagger}\Big)}_{\hat{H}_{\square}}.
    \label{eq:H_QLM}
\end{align}
$\hat{H}_{\kappa}$ describes the minimal coupling between the hardcore bosonic matter fields with the ladder operators $\hat{\phi}_{\mathbf{j}}, \hat{\phi}^{\dag}_{\mathbf{j}}$ located at lattice sites $\mathbf{j} = (j_x, j_y)^\intercal$, mediated by the gauge field spin-$\frac{1}{2}$ operator $\hat{S}^{+}_{\mathbf{j},\mathbf{e}_{\mu}}$ residing on the links connecting adjacent lattice sites $\mathbf{j}$ and $\mathbf{j}+\mathbf{e}_{\mu}$, with $\mathbf{e}_\mu$ the unit vector along the direction $\mu=x,y$. The staggering coefficient $s_{\mathbf{j},\mathbf{e}_{\mu}}$ is direction dependent, with $s_{\mathbf{j},\mathbf{e}_{x}} = +1$, $s_{\mathbf{j},\mathbf{e}_{y}} = (-1)^{j_x}$. $\hat{H}_m$ assigns a staggered mass $m$ for the matter fields, where $s_{\mathbf{j}}$ is defined as $s_{\mathbf{j}} = (-1)^{j_x + j_y}$ such that a particle residing on an even site ($s_{\mathbf{j}} = +1$) denotes a positron with a positive charge, and the absence of a particle on an odd site ($s_{\mathbf{j}} = -1$) corresponds to an electron with a negative charge. $\hat{H}_E$ represents a linear electric field term with strength $g$, which is introduced to generate string tension in the model, since the intrinsic electric field energy $\propto\big(\hat{S}^z_{\mathbf{j},\mathbf{e}_{\mu}}\big)^2{=}\frac{\hat{\mathds{1}}}{4}$ is an energetic constant in this setup owing to the spin-$\frac{1}{2}$ truncation of the electric field. Finally, $\hat{H}_{\square}$, involving the plaquette operator $\hat{U}_{\square}=\hat{S}^{+}_{\mathbf{j},\mathbf{e}_x}\hat{S}^{+}_{\mathbf{j}+\mathbf{e}_x,\mathbf{e}_y}
\hat{S}^{-}_{\mathbf{j}+\mathbf{e}_y,\mathbf{e}_x}
\hat{S}^{-}_{\mathbf{j},\mathbf{e}_y}$, governs the magnetic interactions of the gauge fields with
strength $J$.

The Hamiltonian \eqref{eq:H_QLM} is invariant under local transformations generated by $\hat{G}_{\mathbf{j}} =
\hat{\phi}^{\dagger}_{\mathbf{j}} \hat{\phi}_{\mathbf{j}}
-\frac{1-(-1)^{j_x + j_y}}{2}
-\sum_{\mu}\big(
\hat{S}^z_{\mathbf{j},\mathbf{e}_{\mu}}
-
\hat{S}^z_{\mathbf{j}-\mathbf{e}_{\mu},\mathbf{e}_{\mu}}
\big)$,
which corresponds to a discretized version of Gauss’s
law. Here, we work in the physical sector, i.e., we consider only the gauge-invariant states $\ket{\psi}$ satisfying $\hat{G}_\mathbf{r}\ket{\psi} = g_\mathbf{j}\ket{\psi}$, where $g_\mathbf{j}=0$ for all sites except two: one even lattice site $\mathbf{j}_\text{e}$ and one odd site $\mathbf{j}_\text{o}$ at opposite corners of the lattice that are chosen to host static charges, which we enforce by explicitly setting $g_{\mathbf{j}_\mathrm{e}}=-1$ and $g_{\mathbf{j}_\mathrm{o}}=+1$; see Fig.~\ref{fig:Schematic}\textbf{a}. With this choice, gauge-invariant configurations include electric strings connecting the two static charges. Throughout this work, we focus solely on systems with open boundary conditions.

\textbf{Quantum circuit.---}
We now describe the quantum circuit realization of the system’s time evolution on the \texttt{Quantinuum System Model H2} device, governed by the Hamiltonian \eqref{eq:H_QLM}. We encode matter sites as qubits such that an empty (occupied) site corresponds to $\ket{0}$ ($\ket{1}$), while the gauge fields are encoded with spin-down (spin-up) orientation as $\ket{1}$ ($\ket{0}$). For implementing the time evolution operator, we employ a first-order Trotter decomposition,
\begin{align}
\mathcal{U}(\Delta t) \approx e^{-i \hat{H}_{\kappa} \Delta t} \, e^{-i \hat{H}_{E} \Delta t} \, e^{-i \hat{H}_{m} \Delta t} \, e^{-i \hat{H}_{\square} \Delta t},
\end{align}
where, rather than using a straightforward Trotter decomposition,  we adopt a structured one.
The full circuit for each Trotter step is decomposed into five sublayers arranged such that, within each sublayer, the constituent terms of the Hamiltonian mutually commute and can therefore be implemented in parallel; see Fig.~\ref{fig:Schematic}\textbf{b}. This construction ensures that all the contributions other than the $e^{-i \hat{H}_{m} \Delta t}$ circuit are executed across the first four sublayers---each with a maximum two-qubit gate depth of $7$, as set by the $e^{-i \hat{H}_{\kappa} \Delta t}$ circuit---which collectively realize $e^{-i \hat{H}_{E} \Delta t}$ with local $R_z$ gates, and $e^{-i \hat{H}_{\kappa} \Delta t}$ and $e^{-i \hat{H}_{\square} \Delta t}$ terms through the circuit constructions shown in Fig.~\ref{fig:Schematic}\textbf{c}. The $e^{-i \hat{H}_{m} \Delta t}$ term is implemented in the final sublayer using local $R_z$ gates acting on all matter qubits. Therefore, a single Trotter step always has a two-qubit gate depth of $28$.

\textbf{Genuine $2+1$D string breaking dynamics.---}
To observe clear and physically meaningful string-breaking dynamics, it is essential to operate in the confined phase, where undesired matter creation outside the string is strongly suppressed. This regime is realized when $2m + g \gg \kappa$, since $2m+g$ is the energy cost required to create an electron--positron pair from the vacuum. Dynamically, string breaking occurs when the energy stored in the string equals that to produce an electron--positron pair. In this picture, the energy stored in a string segment made up of $\ell$ links is $\ell\frac{g}{2}$, where $\frac{g}{2}$ is the energy contribution of each link. Breaking such a string segment is thus an order-$\ell$ process, as it requires flipping $\ell$ links and generating an electron--positron pair at the endpoints of the segment. The resulting energy of the broken segment is $2m - \ell \frac{g}{2}$. By equating this to the energy of the original unbroken segment, we obtain the resonance condition $2m = \ell g$. Notably, within the staggered particle framework, the number of links in any breakable string segment must always be odd. We first analyze the on-resonance case, where string breaking is energetically allowed and gives rise to rich $2+1$D dynamics in the presence of a plaquette term, while, in its absence, the evolution remains restricted to effectively $1+1$D dynamics. 

\begin{figure*}
    \centering
    \includegraphics[width = \linewidth]{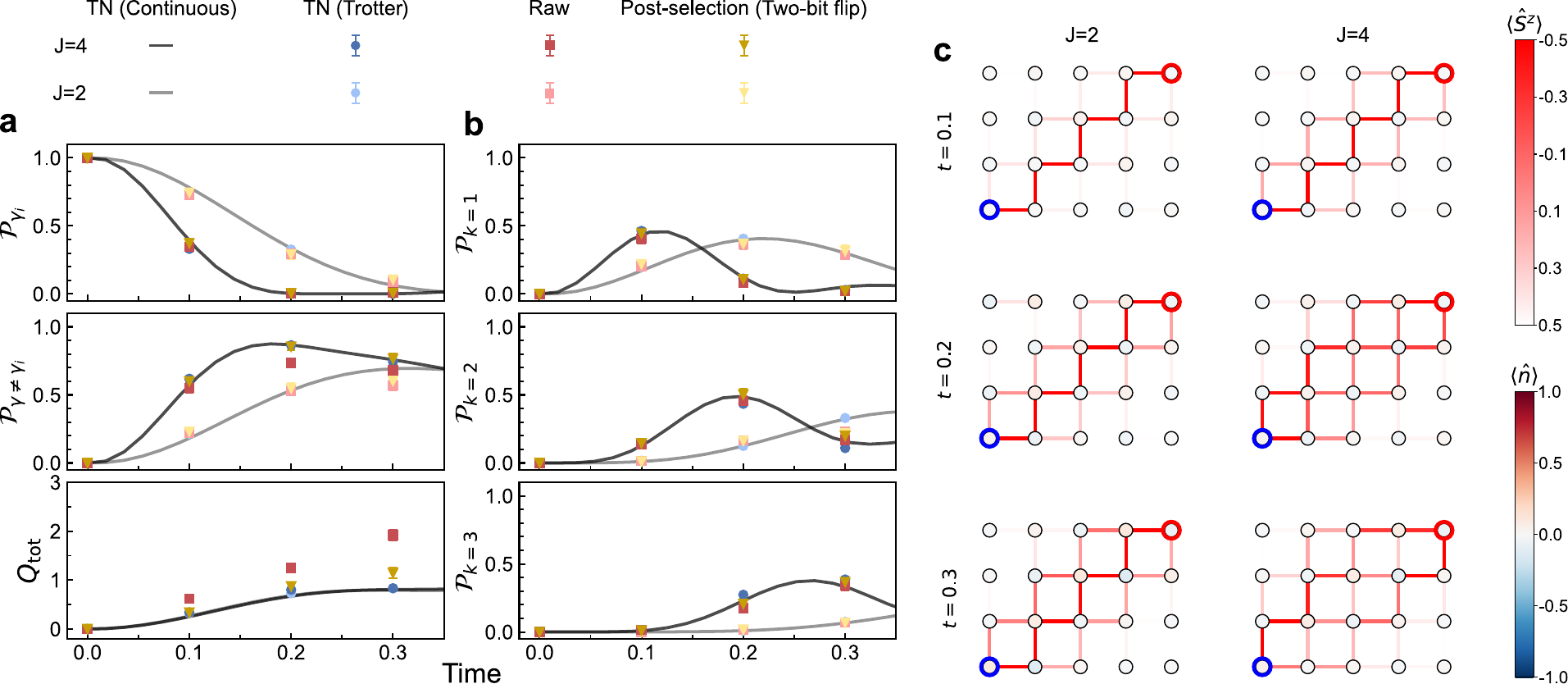}
    \caption{\textbf{Off-resonant $2+1$D string oscillations}. Off-resonant quench of a diagonal string on a $5 \times 4$ lattice with parameters $\kappa =  1$, $m = 5$, $g = 5$, and $J = 2, 4$. $\textbf{\textsf{a}}$ Probability of the initial string configuration, $\mathcal{P}_{\gamma_\mathrm{i}}$, and all other string configurations combined, $\mathcal{P}_{\gamma \neq \gamma_\mathrm{i}}$, as well as the total number of proliferated dynamical charges of the whole system $Q_{\mathrm{tot}}$. We compare both raw and two-bit flip post-selected hardware data against continuous TN simulations and TN simulations of the Trotterized circuit using the \texttt{Qiskit} MPS simulator. $\textbf{\textsf{b}}$ Detailed distribution of the different string configurations generated by the local plaquette term acting at first, second, and third order for different values of $J$.  \textbf{\textsf{c}} Comparison of the spatial distribution of the dynamical charge excitation $\langle \hat{n} \rangle$ and the link observable $\langle \hat{S}^z \rangle $ at each Trotter step obtained from the two-bit flip post-selection result for different values of $J$.}
    
    \label{fig:off_resonant_combined}
\end{figure*}

We prepare an initial product state with all matter sites unoccupied (corresponding to a gauge-invariant ground state in the $m\rightarrow\infty$ limit) hosting a “diagonal” string of negative flux connecting two opposite charges located at the ends of the lattice, with positive flux on the remaining links; see Fig.~\ref{fig:Schematic}\textbf{a},\textbf{b}. The system is then evolved using the previously described Trotterized quantum circuit, followed by full measurements of all qubits in the computational basis. We define the projector $\hat{P}_{\gamma}=\prod_{l \in \gamma} \ketbra{\downarrow}_l$ onto the string configuration $\gamma$ between the two static charges, where $l$ denotes a link. We measure the probability $\mathcal{P}_{\gamma_\mathrm{i}}(t) = \bra{\psi(t)} \hat{P}_{\gamma_\mathrm{i}} \ket{\psi(t)}$ of the time-evolved wave function $\ket{\psi(t)}$ to be in the initial string configuration $\gamma_\mathrm{i}$, 
as well as the probabilities $\mathcal{P}_{k} (t) =  \sum_{\gamma \in \Gamma_k} \bra{\psi(t)} \hat{P}_{\gamma} \ket{\psi(t)}$,
where $\Gamma_k$ denotes the set of minimal string configurations reachable from the initial string $\gamma_\mathrm{i}$ by exactly $k$ operations of the plaquette operator. We also consider the total contribution from all strings different from the initial one,
$\mathcal{P}_{\gamma \neq \gamma_\mathrm{i}}(t) =  \sum_{\gamma \neq \gamma_\mathrm{i}} \bra{\psi(t)} \hat{P}_{\gamma} \ket{\psi(t)}$.  For the hardware data, we compute these probabilities associated with different minimal string configurations, as detailed in the Supplemental Material (SM) \cite{SM}. In addition, we monitor the matter occupation (or charge) within a specified region,
$Q_{\mathcal{S}}= \sum_{{\mathbf{j}} \in \mathcal{S}} \expval{\hat{n}_{\mathbf{j}}}$,
where the staggered particle number operator is 
$\hat{n}_{\mathbf{j}} = (-1)^{j_x + j_y} \hat{\phi}_{\mathbf{j}}^\dagger \hat{\phi}_{\mathbf{j}} - \frac{1 - (-1)^{j_x + j_y}}{2}$, and $\mathcal{S}$ denotes the set of matter qubits considered.

We first investigate string dynamics in the confined regime ($2m + g \gg\kappa)$ at resonance $(2m = g)$ on a $5 \times 4$ lattice using 51 qubits with $\kappa = 1$, $m = 3$, $g = 6$, and $J = 2$. Starting from a diagonal string as shown in Fig.~\ref{fig:Schematic}\textbf{a}, we demonstrate---on quantum hardware---the rich dynamics generated by the presence of an explicit plaquette term. To validate our hardware results, we compare Trotterized TN simulation data---obtained using the \texttt{Qiskit} matrix product state (MPS) method with a fixed bond dimension of 800, continuous-time TN simulation data, raw hardware data, and post-selected hardware data based on three post-selection schemes. In the main text, we show only the scheme with the best performance. Other schemes are detailed in the SM \cite{SM}.

As shown in Fig.~\ref{fig:on_resonance_plaq}\textbf{a}, when the plaquette term is activated ($J = 2)$, the initial string fidelity $\mathcal{P}_{\gamma_\mathrm{i}}$ decays from unity, with near-perfect agreement across all datasets. Being at resonance, the dynamics also exhibits prominent matter creation along the initial string configuration, but also away from it, indicating genuine $2+1$D string breaking; see Fig.~\ref{fig:on_resonance_plaq}\textbf{b}, which shows how the charges at the qubits on and off the initial string path, $Q_\mathrm{on}$ and $Q_\mathrm{off}$, respectively, increase over time. Simultaneously, the probability of the wave function occupying other minimal unbroken string configurations increases at early to intermediate times as the plaquette term drives the dynamics into them. At late times, this probability vanishes as all string configurations are broken and replaced by matter. The first-, second-, and third-order processes, shown in Fig.~\ref{fig:on_resonance_plaq}\textbf{c}, reveal a sequential buildup: the first-order strings peak earliest and most strongly at $t \approx 0.2$, while higher-order strings peak progressively later with smaller amplitudes, all vanishing at late times owing to string breaking. The plaquette term thereby opens access to a broader set of string configurations---including both minimal strings and additional broken string configurations---extending the dynamics beyond the subspace of the initial string, which is the hallmark of genuine $2+1$D behavior. Further highlighting the latter is the sequence of snapshots shown in Fig.~\ref{fig:on_resonance_plaq}\textbf{d}, showing the spread of matter and string fluctuations over the entire lattice in time.

In contrast, when the plaquette term is turned off ($J = 0$), the evolution is confined to the subspace spanned by broken and unbroken string configurations along the initial string, leading to string revivals, as illustrated in Fig.~\ref{fig:on_resonance_no_plaq}, and confirming the findings of a previous TN study of the case of no plaquettes \cite{Tian2025RolePlaquetteTerm}. Focusing on a $4\times3$ lattice for this case with $\kappa = 1$, $m = 3$, $g = 6$, and $J=0$, we observe that the string only breaks along its initial orientation with matter creation confined along the initial string and hardly occurring away from it, while the population of strings across all higher orders remains near zero, clearly demonstrating the absence of fully-fledged $2+1$D dynamics.

For both values of $J$, the raw hardware data shows good agreement with the continuous-time TN data and Trotterized \texttt{Qiskit} data, with the post-selection scheme consistently improving upon the raw results.

\textbf{Off-resonant string dynamics.---}
We now set $g\neq2m$, tuning away from resonance while varying $J$ to investigate the effect of the plaquette term on $2+1$D string oscillations. We demonstrate this on quantum hardware for a $5\times4$ lattice for two values of $J = 2,4$, with the remaining parameters fixed at $\kappa = 1$, $m = 5$, and $g = 5$. The results, displayed in Fig.~\ref{fig:off_resonant_combined}, show great agreement between (both raw and post-selected) hardware and (continuous-time and Trotterized) TN data, with the latter performed using the \texttt{Qiskit} MPS method as before.

For both values of $J$, the initial string probability $\mathcal{P}_{\gamma_\mathrm{i}}$ decays rapidly towards zero, while the probability $\mathcal{P}_{\gamma\neq\gamma_\mathrm{i}}$ of occupying other minimal string configurations increases rapidly, as shown in Fig.~\ref{fig:off_resonant_combined}\textbf{a}. This behavior is more pronounced with larger $J$. This is accompanied by matter creation, which is expected since $m$ is not infinite; however, there is no string breaking. Interestingly, the total matter occupation $Q_{\mathrm{tot}}$ remains independent of $J$ within the considered timescale. This is because string dynamics is governed by $J$, while matter creation is induced by the weaker hopping $\kappa$ ($\kappa < J$), and is additionally energetically suppressed. Consequently, string oscillations occur on shorter timescales and dominate the early dynamics, leaving $Q_{\mathrm{tot}}$ insensitive to $J$.
Figure~\ref{fig:off_resonant_combined}\textbf{b} shows the sequential buildup of first-, second-, and third-order minimal string configurations, with their timescales sped up with $J$. The effect of $J$ is further shown in the averaged snapshots in Fig.~\ref{fig:off_resonant_combined}\textbf{c}, where, at any given time, the string shows a greater spread of oscillations for larger $J$.

\textbf{Conclusion and outlook.---}
In this work, we realized a U$(1)$ lattice gauge theory in $2+1$ dimensions with a tunable plaquette interaction on a trapped-ion quantum computer, and used it to probe far-from-equilibrium string dynamics. By directly comparing dynamics with and without the plaquette term, we demonstrated that only in its presence do strings explore the lattice plane and generate matter away from their initial configuration, establishing the onset of genuine $2+1$D dynamics, including multi-order string breaking. In contrast, when the plaquette term is absent, the evolution remains confined to effectively $1+1$D processes despite the higher-dimensional geometry. Our results identify the plaquette interaction as the key ingredient enabling higher-dimensional gauge dynamics and provide a direct experimental validation of its role.

Beyond demonstrating this qualitative change in dynamics, our implementation achieves the largest known digital quantum simulation of a $2+1$D U$(1)$ LGT to date, combining a scalable encoding of matter and gauge fields with an efficient circuit design. The tunability of the plaquette interaction further opens the door to studying dynamical crossovers between effectively $1+1$D and $2+1$D regimes. More broadly, our work establishes a scalable pathway toward probing real-time, nonperturbative dynamics of gauge theories in higher dimensions, toward regimes where classical simulability is strongly challenged.

\medskip
\footnotesize

\textit{Note.---}
In the same \texttt{arXiv} listing as this paper, a parallel submission \cite{Xu2026ObservationOfGlueballExcitations} involving some of the current authors will appear, where \texttt{Quantinuum System Model H2} is used to observe glueball excitations and string breaking in a $2+1$D $\mathbb{Z}_2$ LGT on a $6\times5$ square lattice with a tunable plaquette term.

\medskip

\textbf{Acknowledgments.---}
We thank Umberto Borla, Michael Mills, and Andrew Potter for useful comments on the manuscript.
R.J., Y.T., K.X., N.S.S., J.J.O., and J.C.H.~acknowledge funding by the Max Planck Society, the Deutsche Forschungsgemeinschaft (DFG, German Research Foundation) under Germany’s Excellence Strategy – EXC-2111 – 390814868, and the European Research Council (ERC) under the European Union’s Horizon Europe research and innovation program (Grant Agreement No.~101165667)—ERC Starting Grant QuSiGauge. Views and opinions expressed are, however, those of the author(s) only and do not necessarily reflect those of the European Union or the European Research Council Executive Agency. Neither the European Union nor the granting authority can be held responsible for them.  All experiments were run on \texttt{Quantinuum H2-2} quantum computer, powered by \texttt{Honeywell}. This work is part of the Quantum Computing for High-Energy Physics (QC4HEP) working group.
\normalsize

\bibliography{biblio}

\clearpage
\pagebreak
\setcounter{equation}{0}
\setcounter{figure}{0}
\setcounter{table}{0}
\setcounter{page}{1}
\setcounter{section}{0}
\makeatletter
\renewcommand{\theequation}{S\arabic{equation}}
\renewcommand{\thefigure}{S\arabic{figure}}
\renewcommand{\thesection}{S\Roman{section}}
\renewcommand{\thepage}{\arabic{page}}
\renewcommand{\thetable}{S\arabic{table}}
\vspace{0cm}
%\twocolumngrid
\normalsize

\onecolumngrid
\begin{center}
    \textbf{\large Supplemental Online Material for \\``Observation of genuine $2+1$D string dynamics in a U$(1)$ lattice gauge theory with a tunable plaquette term on a trapped-ion quantum computer''}\\[5pt]
    \vspace{0.1cm}
\end{center}

\section{Experimental Methods}
\subsection{Hardware specifications}
All experiments were performed on the \texttt{Quantinuum System Model H2} system \cite{quantinuum-h2-2}, a trapped-ion quantum processor in which qubits are encoded in the atomic hyperfine states of $^{171}$Yb$^+$ ions. The system provides 56 physical qubits arranged in a Quantum Charge-Coupled Device (QCCD) architecture, comprising two connected linear trap sections with four gate zones. Qubit connectivity is any-to-any, achieved by physically shuttling ions between interaction zones, with up to four two-qubit operations executable in parallel. The native gate set consists of single-qubit rotations and a parameterized-angle two-qubit $ZZ$ gate.
The system additionally supports mid-circuit measurement with conditioned branching and qubit reuse.

Benchmarked gate infidelities for \texttt{Quantinuum H2-2} are $2.8\times10^{-5}$ ($\pm 3.6\times10^{-6}$) for single-qubit gates, $8.4\times10^{-4}$ ($\pm 4.8\times10^{-5}$) for two-qubit gates, and $6.7 \times10^{-4}$ ($\pm 8.7\times10^{-5}$) for state preparation and measurement of $|0\rangle$.
Memory error per depth-1 circuit time is $1.2\times10^{-4}$ ($\pm 2.0\times10^{-5}$), and measurement cross-talk error is $2.2\times10^{-5}$ ($\pm 5.3\times10^{-7}$).
Reported infidelities are averaged over all operational zones and are saved on GitHub with all the relevant randomized benchmarking data \cite{quantinuum-performance}.

\subsection{Dynamical Decoupling}
Between gate operations, qubits experience memory errors due to both coherent and incoherent dephasing, as well as leakage. To suppress coherent dephasing, we employ dynamical decoupling (DD), which inserts periodic $X$ pulses during idle periods to approximately cancel unwanted phase accumulation. 

Since the ions are dynamically routed during circuit execution, the exact idle time of each qubit cannot be determined \textit{a priori}. Consequently, DD sequences are scheduled at the compiler level (a capability supported on the \texttt{Quantinuum System Model H2} device), using a threshold of about 0.03 s to ensure effective suppression of dephasing during inactive periods. Pulse insertion is performed automatically based on estimated idle times.

In addition, several compiler-level optimizations are employed to improve DD efficiency and reduce overhead. Rotation waveforms are merged whenever possible to shorten overall circuit duration, and DD pulses are incorporated into existing single-qubit operations when feasible.

\subsection{Resource analysis}
The single-qubit gate count, two-qubit gate count, two-qubit gate depth, and 
total depth per Trotter step are given in Table~\ref{tab:resource_estimates}.

\begin{table}[h!]
\centering
\begin{tabular}{|c|c|c|c|c|c|}
\hline
System size & Qubits used & 1q gate count & 2q gate count & 2q gate depth & Total depth \\ \hline
5 $\times$ 4 ($J \neq 0$) & 51 & 682 & 385 & 28 & 159 \\ \hline
4 $\times$ 3 ($J = 0$) & 29 & 222 & 119 & 28 & 137 \\ \hline
\end{tabular}
\caption{Resource requirements per single Trotter step for different lattice systems.}
\label{tab:resource_estimates}
\end{table}

\section{Measuring string probabilities using local projectors}

The probability (or fidelity) of each string configuration is computed by applying a local projector onto the relevant link qubits. Specifically, for a given string configuration $\gamma$, we define the projector
\begin{equation}
    \hat{P}_{\gamma} = \prod_{\ell \in \gamma} \hat{P}_\ell^{(1)},
\end{equation}
where the product runs over all link qubits $\ell$ belonging to the configuration $\gamma$, and $\hat{P}_\ell^{(1)} = \ket{1}\bra{1}_\ell$ projects link $\ell$ onto the state $\ket{1}$. The string probability is then evaluated as the expectation value

\begin{equation}
    p_{\gamma}(t) = \frac{1}{N_{\mathrm{shots}}} \sum_{i=1}^{N_{\mathrm{shots}}} \prod_{\ell \in \gamma} b_\ell^{(i)},
\end{equation}
where $N_{\mathrm{shots}}$ is the total number of measurement shots and $b_{\ell}^{(i)} \in \{0,1\}$ is the eigenvalue of $\hat{P}_{\ell}^{i}$ for the $i^\mathrm{th}$ shot. This quantity measures the probability that all link qubits along the string configuration $\gamma$ are simultaneously in state $\ket{1}$, irrespective of the state of the remaining qubits not belonging to $\gamma$. For all the simulations, we set $N_{\text{shots}} = 200$.

\section{Comparison between the local projector and the global projector}
\begin{figure*}
        \includegraphics[width=\linewidth]{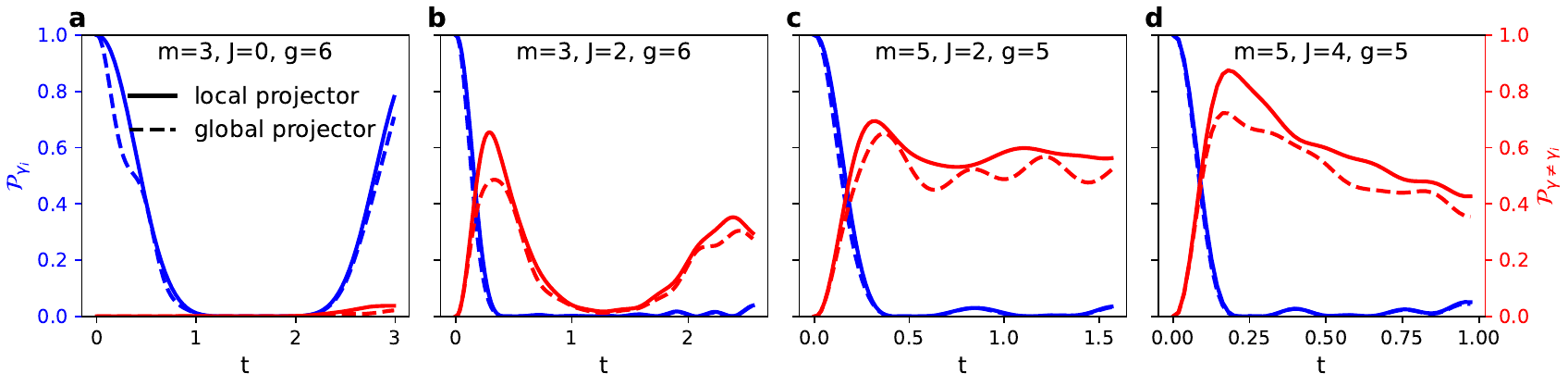}
        \caption{String probabilities measured using local (solid lines) and global (dashed lines) projectors. The initial string is shown in blue, while other string configurations are shown in red. Four different parameter regimes, corresponding to those simulated in the main text, are presented in panels \textbf{\textsf{a}}–\textbf{\textsf{d}}. The result is obtained via continuous-time TN simulations with bond dimensions up to 400.}
        \label{fig:compare_local_global}
\end{figure*}

In this section, we confirm that the system remains in the confined regime by comparing the string probability extracted from both local and global measures. Beyond the local projector $\hat{P}_\gamma$, we also evaluate the overlap between the time-evolved state and the state containing only a string excitation on top of the vacuum  $\langle{\psi_{\gamma}}|{\psi(t)}\rangle$. Unlike the local projector, this global measure excludes states in which the string excitation is accompanied by additional matter–antimatter pairs created elsewhere in the patch. Such a contribution is also counted by the local projector. In the confined regime, the difference between these two measures is expected to remain small, reflecting the suppression of matter creation. In the limiting case of infinite $m$ and $g$, the two measures become identical. In Fig.~\ref{fig:compare_local_global} we compare them across the full parameter range explored in the main text, for both the fidelity to the initial string state and the probability of populating other string configurations. Throughout the timescales considered here, the discrepancy between the two remains small.

\iffalse
\begin{figure}[H]
        \centering
        \includegraphics[width=0.6\linewidth]{Layer_1.pdf}
        \includegraphics[width=0.6\linewidth]{Layer_3.pdf}
        \includegraphics[width=0.6\linewidth]{Layer_2.pdf}
        \includegraphics[width=0.6\linewidth]{Layer_4.pdf}
        \includegraphics[width=0.6\linewidth]{Layer_5.pdf}
        
        \caption{Trotter Circuit decomposition -  One Trotter step is partitioned into four steps as shown in the figure, where the hopping term is implemented for the highlighted links and the plaquette term is implemented for the links in the diamond. The hopping terms for each step commute with each other and the plaquette term, therefore the two-qubit gate depth of the subcircuit is 7 (depth of the plaquette term), bringing the total two-qubit gate depth of one Trotter step to 28.}
        \label{fig:circ_decomp}
    \end{figure}

\fi

\section{Post-selection schemes}
We employ three post-selection schemes as error mitigation strategies for our noisy hardware data. In the first approach, \textit{hard post-selection}, we discard all bitstrings that correspond to configurations not dynamically accessible within our lattice under the plaquette and hopping terms.

To determine the set of bitstrings dynamically reachable from the chosen initial product state, we perform a breadth-first search in configuration space. Starting from the initial state, we iteratively apply all allowed local off-diagonal operators, namely the plaquette and hopping terms. Whenever such an operation generates a new bitstring, it is added to the queue and recorded as reachable. Since each configuration is visited only once, the procedure terminates when the queue is empty, yielding the full dynamically connected sector associated with the initial state. Owing to Hilbert-space fragmentation, this sector is much smaller than the gauge-invariant Hilbert space obtained by imposing Gauss's law alone. For the $5\times4$ system considered here, the effective Hilbert-space dimension is $636{,}487$, compared with the full dimension $2^{51}$. 

For a total of $N$ measurement shots, the probability of a given state $\ket{\psi}$ at the $i^{\text{th}}$ Trotter step is defined as
\begin{equation}
p_{\psi}(i) = \frac{N^{\text{post}}_{\psi}(i)}{N^{\text{post}}(i)},
\end{equation}
where $N^{\text{post}}_{\psi}(i)$ is the number of occurrences of the bitstring corresponding to $\ket{\psi}$ after post-selection, and $N^{\text{post}}(i)$ is the total number of retained (post-selected) counts. This method strictly enforces the physical constraints of the model, ensuring that only valid configurations contribute to the measured observables.

In the second approach, \textit{soft post-selection}, we relax this constraint by additionally including bitstrings that can be converted into a valid configuration through one- or two-bit flip(s) anywhere in the string. In practice, this method increases the effective statistics compared to hard post-selection, as more measurement outcomes are retained. At the same time, it preserves the essential structure of the dynamics, since only configurations that are one- or two-bit flip(s) away from valid states are included. As a result, soft post-selection provides a compromise between noise suppression and data retention, preserving the essential features of the dynamics while mitigating the impact of local errors.

\begin{figure*}
        \includegraphics[width=\linewidth]{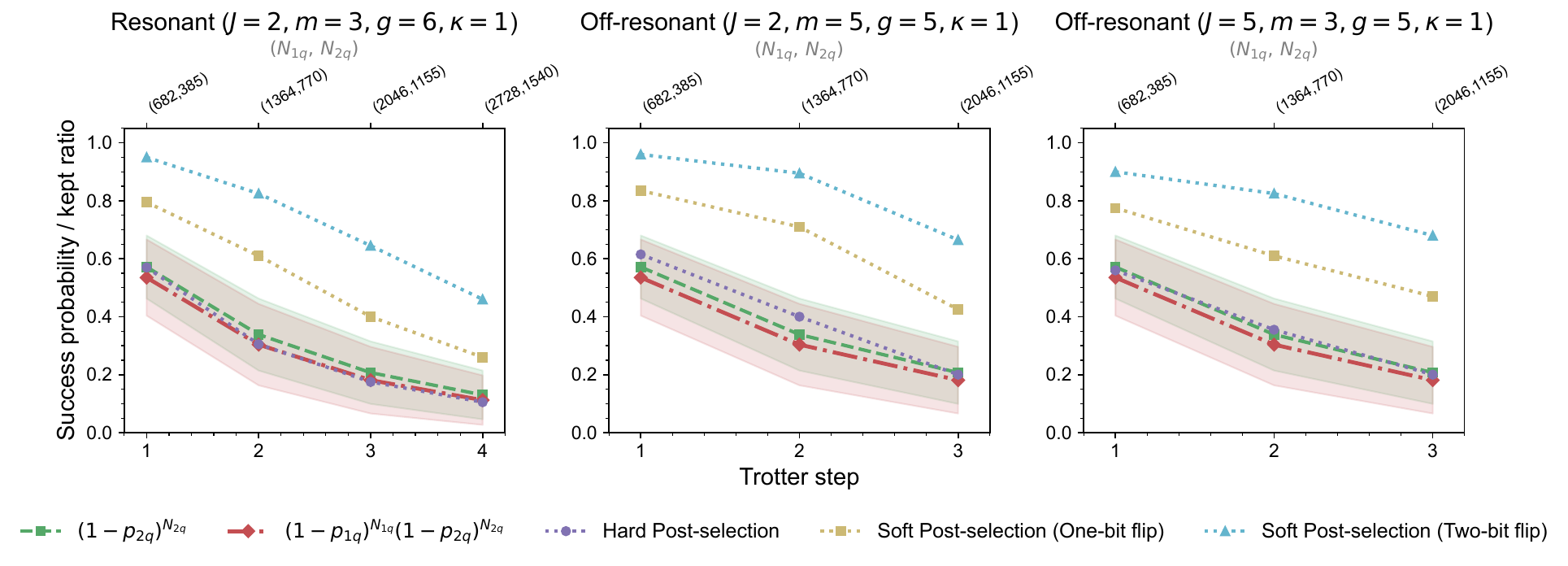}
        \caption{The expected success probability averaged over typical and max noise, and post-selection kept ratios at different Trotter steps, employed for the different parameter regimes studied on the $5\times4$ system in the main text. $N_{1q}$ and $N_{2q}$ refer to the number of single and two-qubit gates used, respectively. We choose $p_{1q} = 3 \times 10^{-5}$ $(2 \times 10^{-4})$, and $p_{2q} = 1 \times 10^{-3}$ $(2 \times 10^{-3})$ for typical (maximum) single and two-qubit gate infidelities.}
        \label{fig:post_ratio}
    \end{figure*}

The expected success probability of obtaining a valid bitstring due to single- and two-qubit gate infidelities, along with the post-selection retention ratios for different schemes, are shown in Fig.~\ref{fig:post_ratio}. The retention ratio for hard post-selection lies within the bounds defined by maximal and typical noise, while soft post-selection schemes yield higher retention ratios.

\begin{figure*}
        \includegraphics[width=\linewidth]{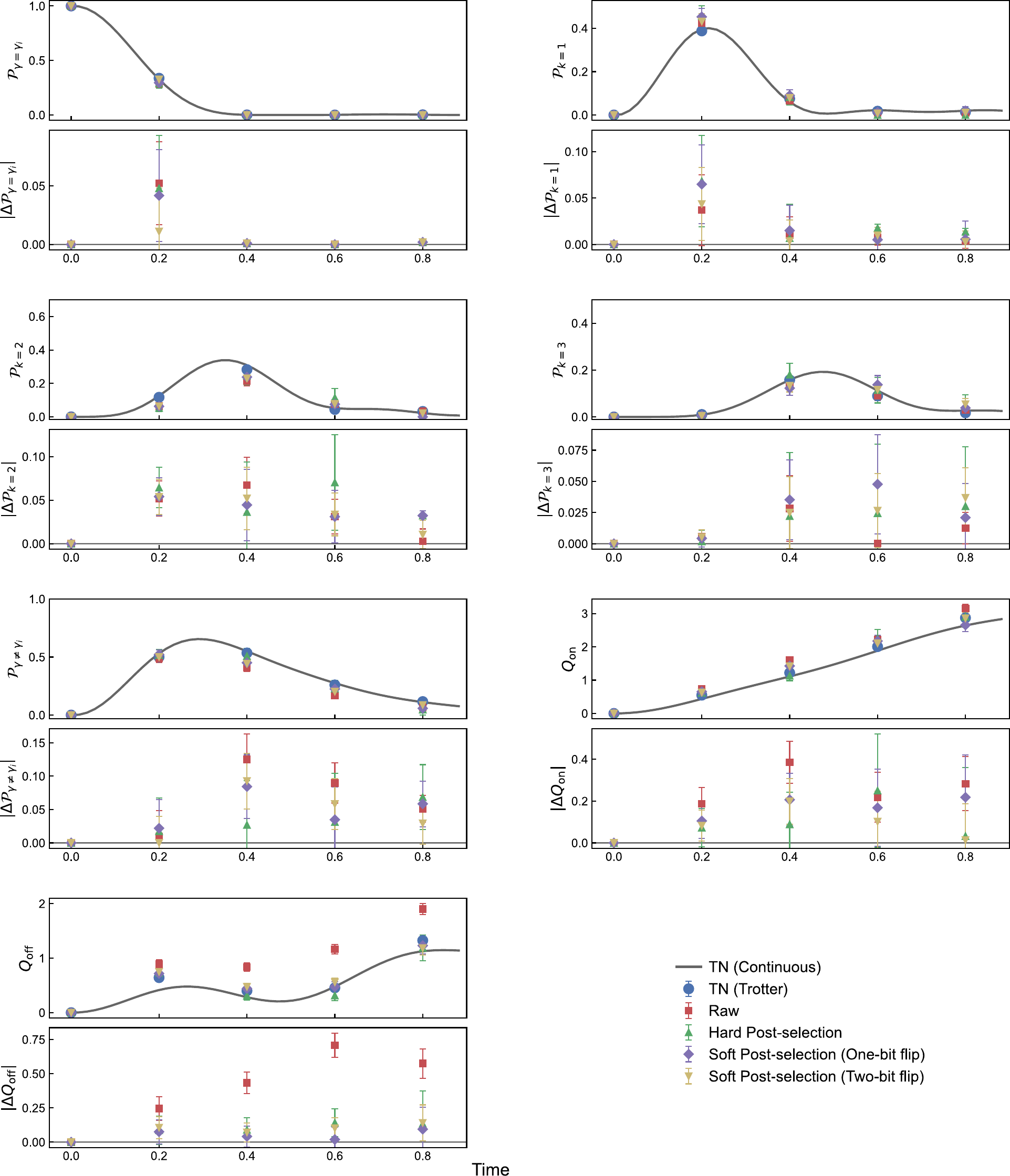}
        \caption{Comparison of different post-selection schemes for an on-resonant quench of the diagonal string on a $5\times4$ lattice with the parameters $\kappa = 1$, $m = 3$, $g = 6$, and $J = 2$. The lower panels for each observable show the absolute deviation of each dataset from the Trotterized TN data.}
        \label{fig:all_J_2}
    \end{figure*}

We also show plots in Fig.~\ref{fig:all_J_2} to Fig.~\ref{fig:tomography_J=4_off}   comparing the performance of these schemes for all parameter regimes discussed in the main text. For an observable \(X\), we define \( |\Delta X(t)| = \left| X_{\mathrm{method}}(t) - X_{\mathrm{TN\,(Trotter)}}(t) \right|\), where \(X_{\mathrm{method}}(t)\) denotes the value obtained from a given hardware-processing method (raw, hard post-selection, one-bit flip, or two-bit flip post-selection), and \(X_{\mathrm{TN\,(Trotter)}}(t)\) is the corresponding noiseless Trotterized result. The uncertainty of the deviation is obtained via standard error propagation assuming independent errors, \(\sigma_{\Delta X}(t) = \sqrt{ \sigma_{\mathrm{method}}(t)^2 + \sigma_{\mathrm{TN\,(Trotter)}}(t)^2 }\).

\section{Numerical details}

\begin{figure*}
        \includegraphics[width=\linewidth]{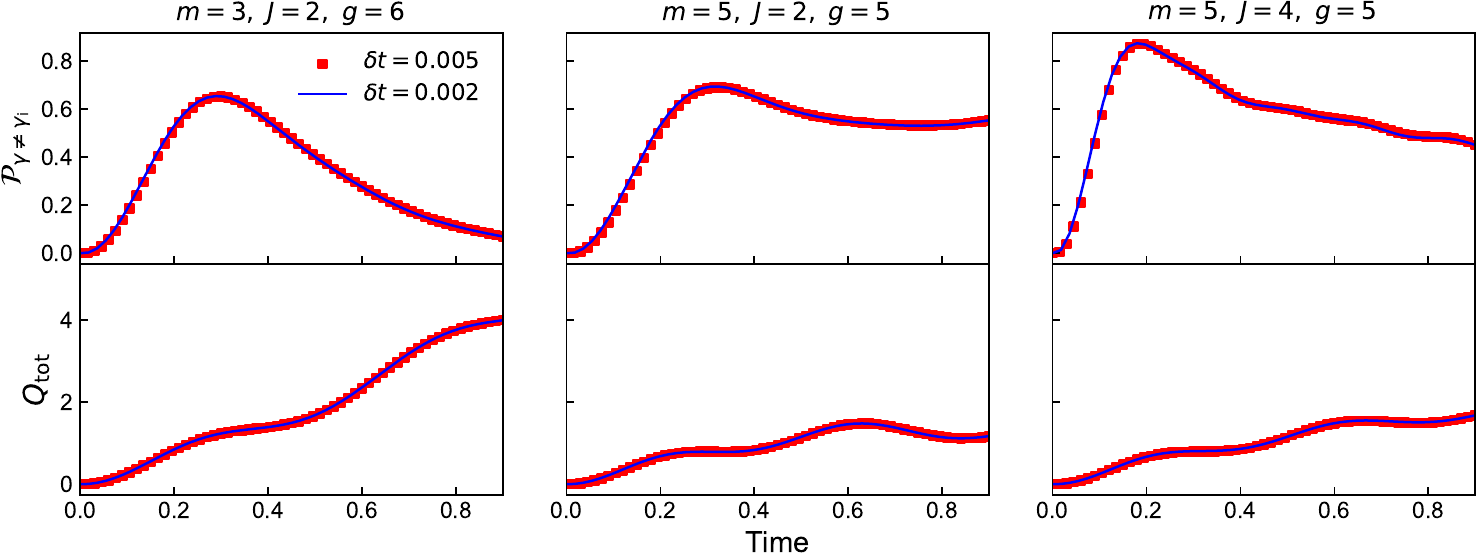}
        \caption{The convergence with respect to the time step $\delta t$ of the total charge $Q_{\mathrm{tot}}$ and the probability of occupying minimal string configurations other than the initial one, $\mathcal{P}_{\gamma \neq \gamma_{\mathrm{i}}}$, evaluated across the different parameter regimes discussed in the main text. The algorithm incorporates controlled bond expansion, enabling a dynamic increase of the bond dimension during unitary time evolution. For the parameters considered in the main text, the bond dimension is increased up to $500$.}
        \label{fig:numerical}
    \end{figure*}

All continuous data used to benchmark the quantum circuit for the $5\times4$ system are obtained using MPS numerics---namely, the time-dependent variational principle (TDVP) algorithm \cite{Haegeman2016UnifyingTimeEvolution} implemented in the \texttt{Matrix Product Toolkit} \cite{mptoolkit} with a time step $\delta t = 0.002/\kappa$. The algorithm employs controlled bond expansion techniques, with RSVD-based pre- and post-expansion \cite{McCulloch2024CommentControlledBond}, dynamically increasing the bond dimension throughout the unitary time evolution. For the parameters considered in the main text, the bond dimension is increased up to $500$, with a truncation tolerance of $10^{-14}$. Convergence with respect to the time step is shown in Fig.~\ref{fig:numerical}. For the simulations for the $4 \times 3$ system, we used exact diagonalization.

\begin{figure*}
        \includegraphics[width=\linewidth]{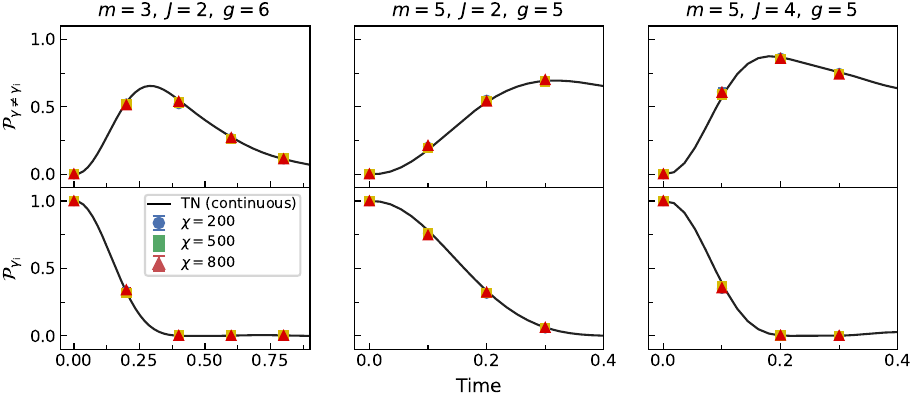}
        \caption{The convergence test of the noiseless circuit simulation of the $5 \times 4$ system using the \texttt{Qiskit} MPS simulator with respect to the bond dimension $\chi$ of the probability of populating minimal string configurations other than the initial one, $\mathcal{P}_{\gamma \neq \gamma_{\mathrm{i}}}$, and remaining in the initial one, $\mathcal{P}_{\gamma_{\mathrm{i}}}$, evaluated across the different parameter regimes discussed in the main text.}
        \label{fig:numerical_circ}
    \end{figure*}

For the noiseless circuit simulations of the $5 \times 4$ system, we employed the MPS circuit simulator method available in \texttt{Qiskit}. To verify convergence with respect to the bond dimension $\chi$, we repeated the simulations for $\chi \in \{200, 500, 800\}$ across all the different parameter regimes discussed in the main text. Each simulation was averaged over five independent runs, and the results are indistinguishable across all values of $\chi$, confirming convergence as shown in Fig.~\ref{fig:numerical_circ}. For the $4 \times 3$ system, we used the \texttt{Qiskit statevector} method, for which no such convergence study is required.

\begin{figure*}
        \includegraphics[width=\linewidth]{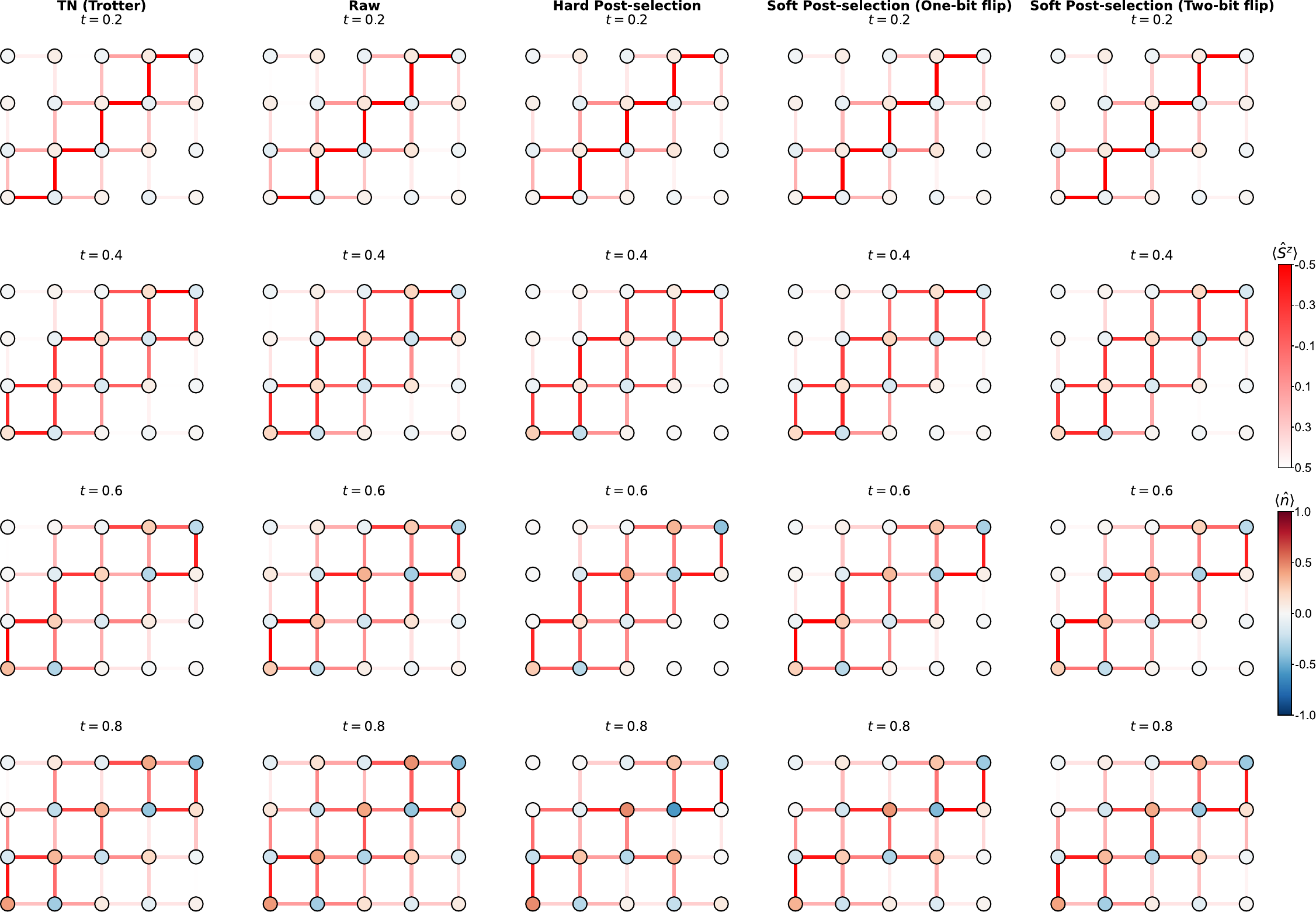}
        \caption{Snapshots of the on-resonant quench dynamics of an initial diagonal string on a $5\times4$ lattice in the presence of a plaquette term comparing the raw hardware and different post-selected data with the Trotterized TN data.}
        \label{fig:tomography_J=2}
    \end{figure*}

 \begin{figure*}
        \includegraphics[width=\linewidth]{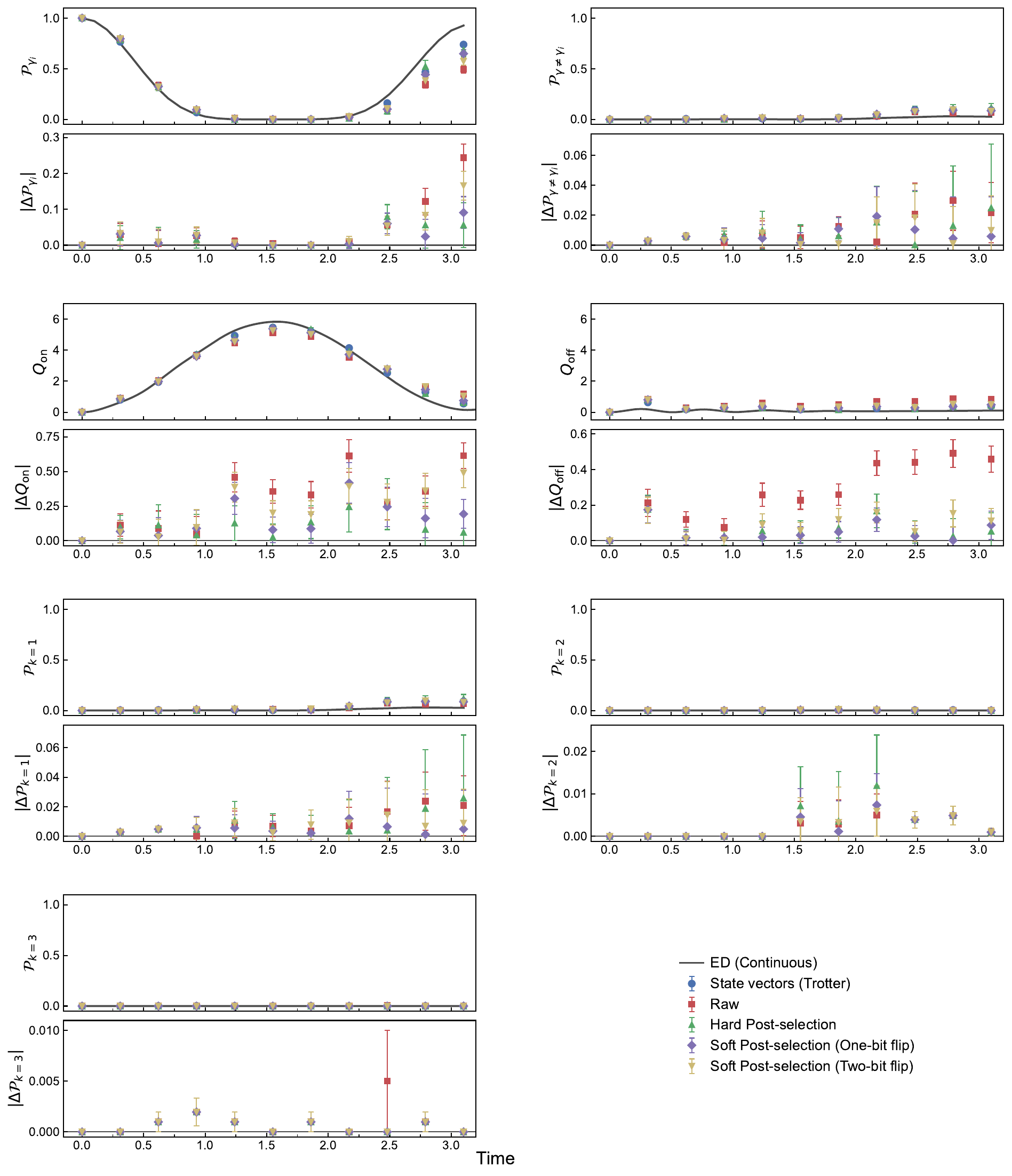}
        \caption{Comparison of different post-selection schemes for an on-resonant quench of the diagonal string on a $4 \times 3$ lattice with the parameters $\kappa = 1$, $m = 3$, $g = 6$, and $J = 0$.  The lower panels for each observable show the absolute deviation of each dataset from the Trotterized TN data. }
        \label{fig:all_J_0}
    \end{figure*}

\begin{figure*}
        \includegraphics[width=\linewidth]{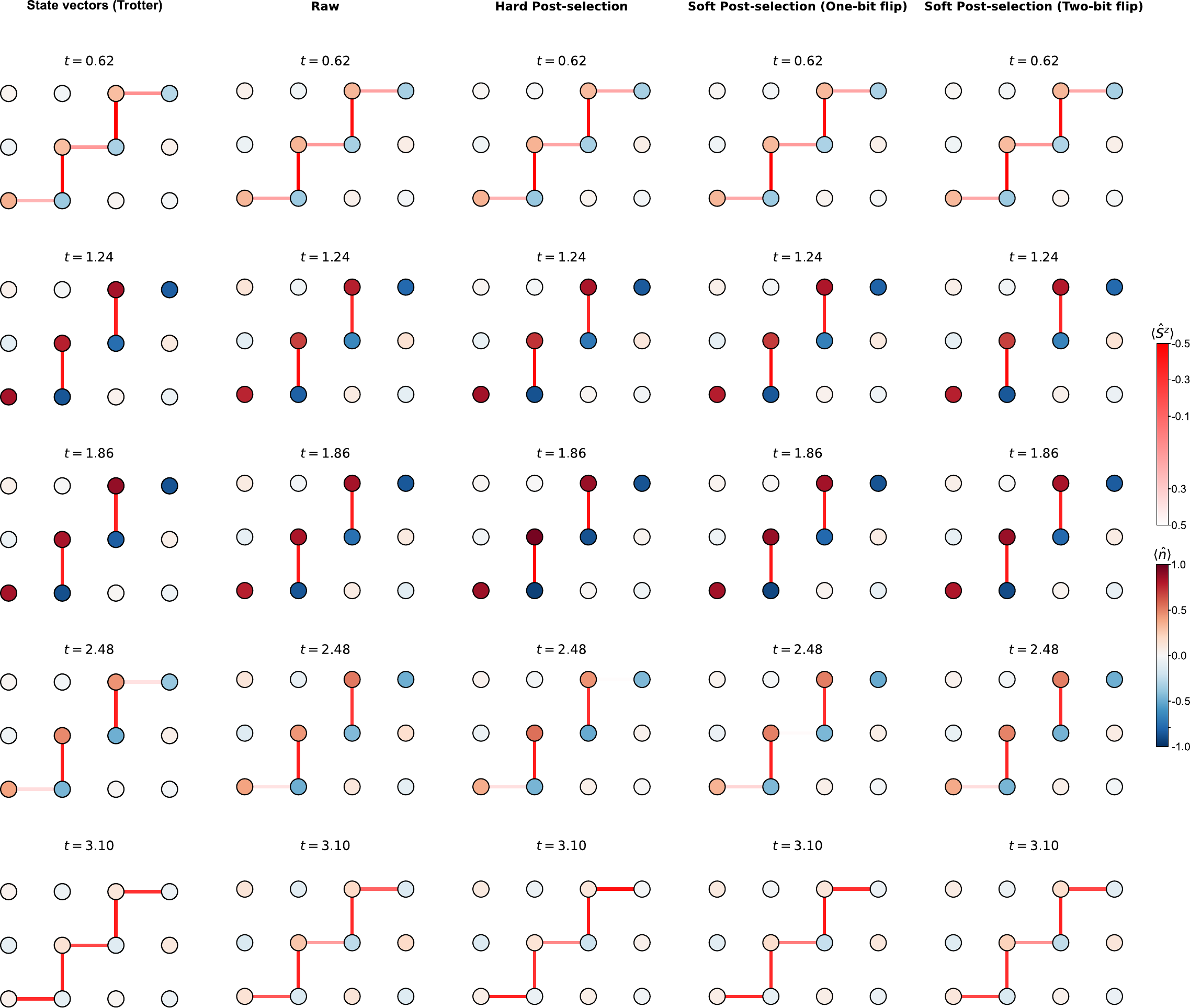}
        \caption{Snapshots of the on-resonant quench dynamics of an initial diagonal string on a $4\times3$ lattice comparing the raw hardware and different post-selected data with the Trotterized TN data.}
        \label{fig:tomography_J=0_onresonance}
    \end{figure*}

\begin{figure*}
        \includegraphics[width=\linewidth]{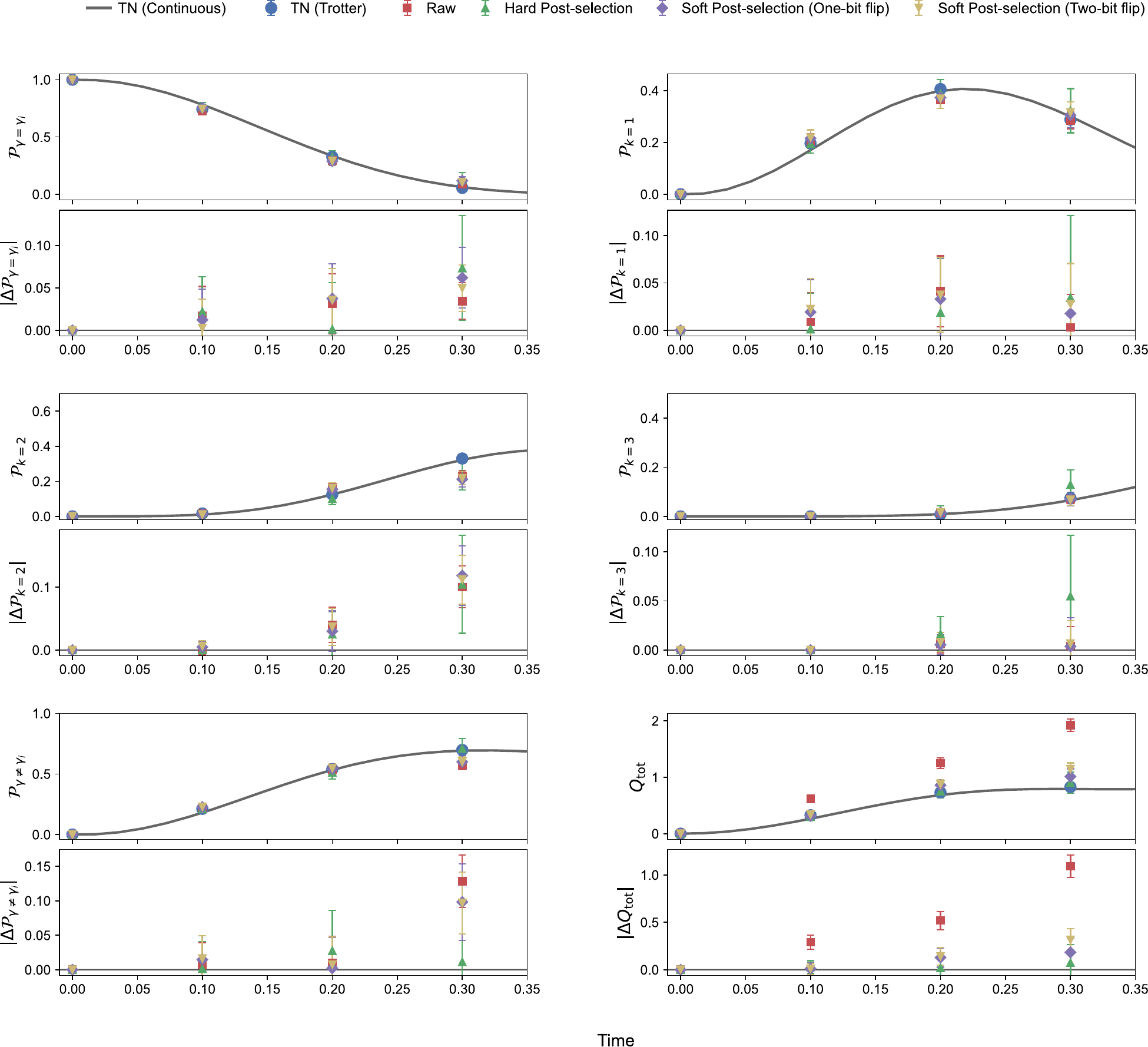}
        \caption{Comparison of different post-selection schemes for an on-resonant quench of the diagonal string on a $5\times4$ lattice with the parameters $\kappa = 1$, $m = 5$, $g = 5$, and $J = 2$.  The lower panels for each observable show the absolute deviation of each dataset from the Trotterized TN data.}
        \label{fig:tomography_J=2_off}
    \end{figure*}

\begin{figure*}
        \includegraphics[width=\linewidth]{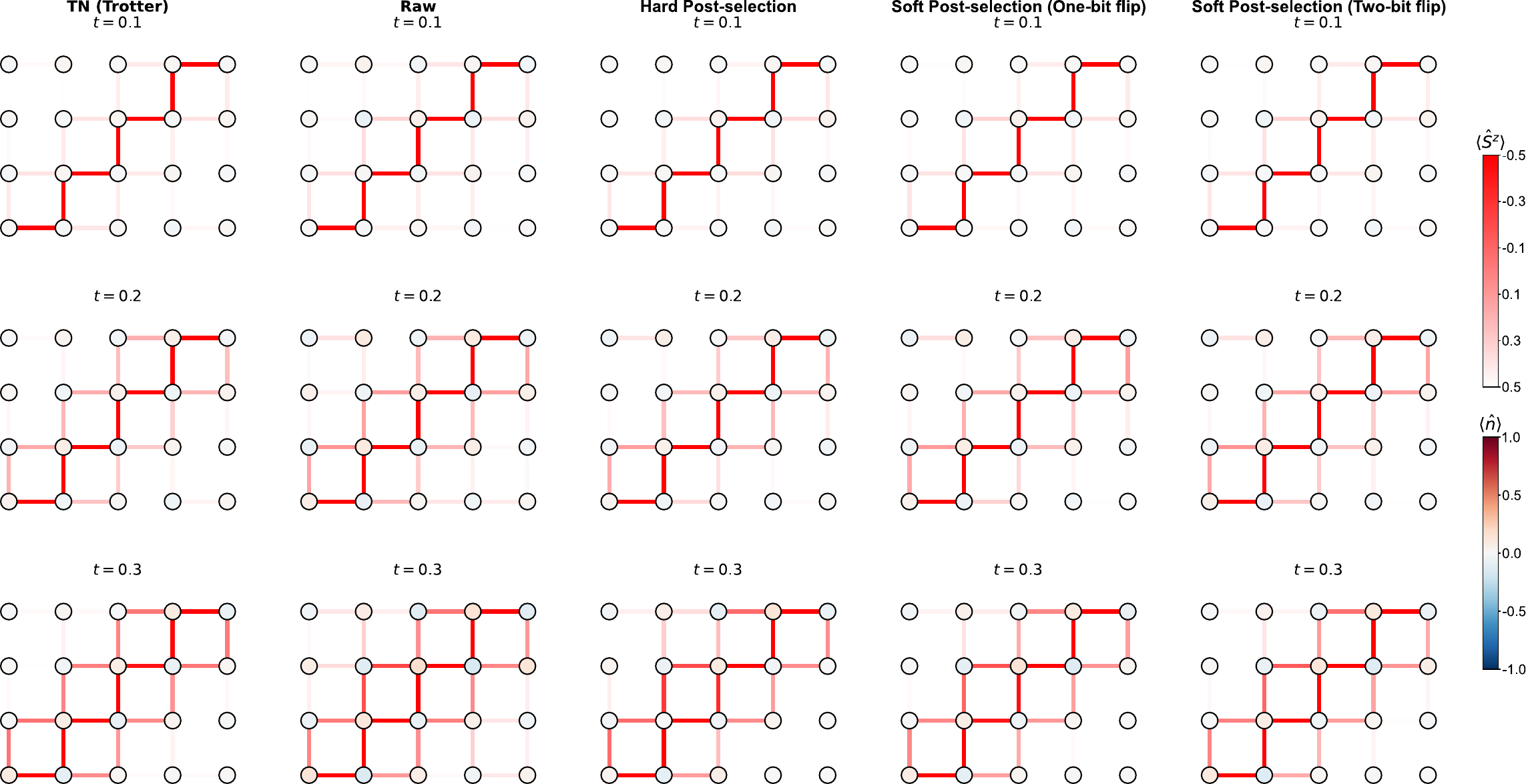}
        \caption{Snapshots of the off-resonant quench dynamics of an initial diagonal string on a $ 5\times4 $ lattice in the presence of a plaquette term with $J = 2$ comparing the raw hardware and different post-selected data with the Trotterized TN data.}
        \label{fig:tomography_J=2_off}
    \end{figure*}

\begin{figure*}
        \includegraphics[width=\linewidth]{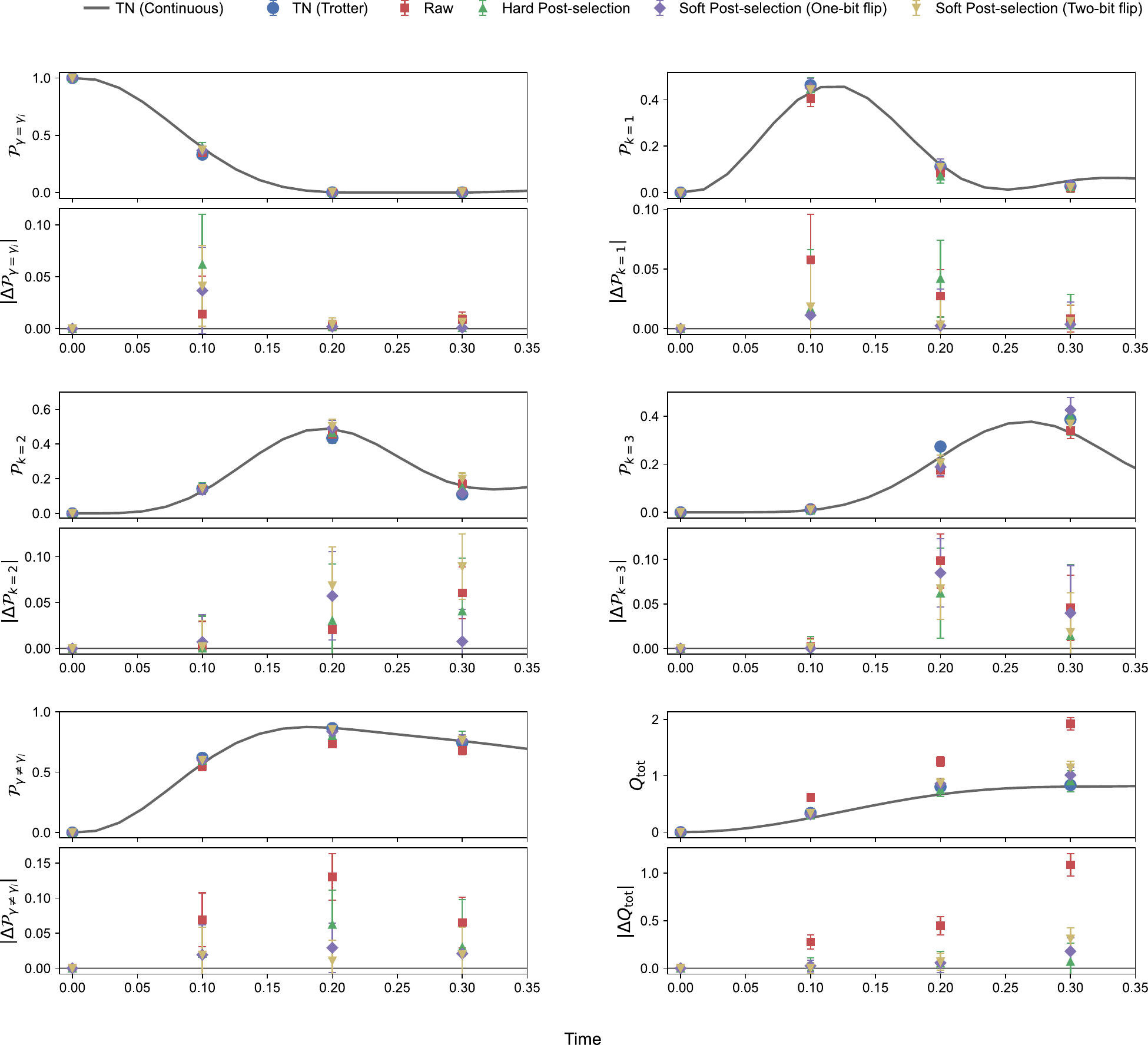}
        \caption{Comparison of different post-selection schemes for an on-resonant quench of the diagonal string on a $5\times4$ lattice with the parameters $\kappa = 1$, $m = 5$, $g = 5$, and $J = 4$.  The lower panels for each observable show the absolute deviation of each dataset from the Trotterized TN data.}
        \label{fig:tomography_J=4_off}
    \end{figure*}

\begin{figure*}
        \includegraphics[width=\linewidth]{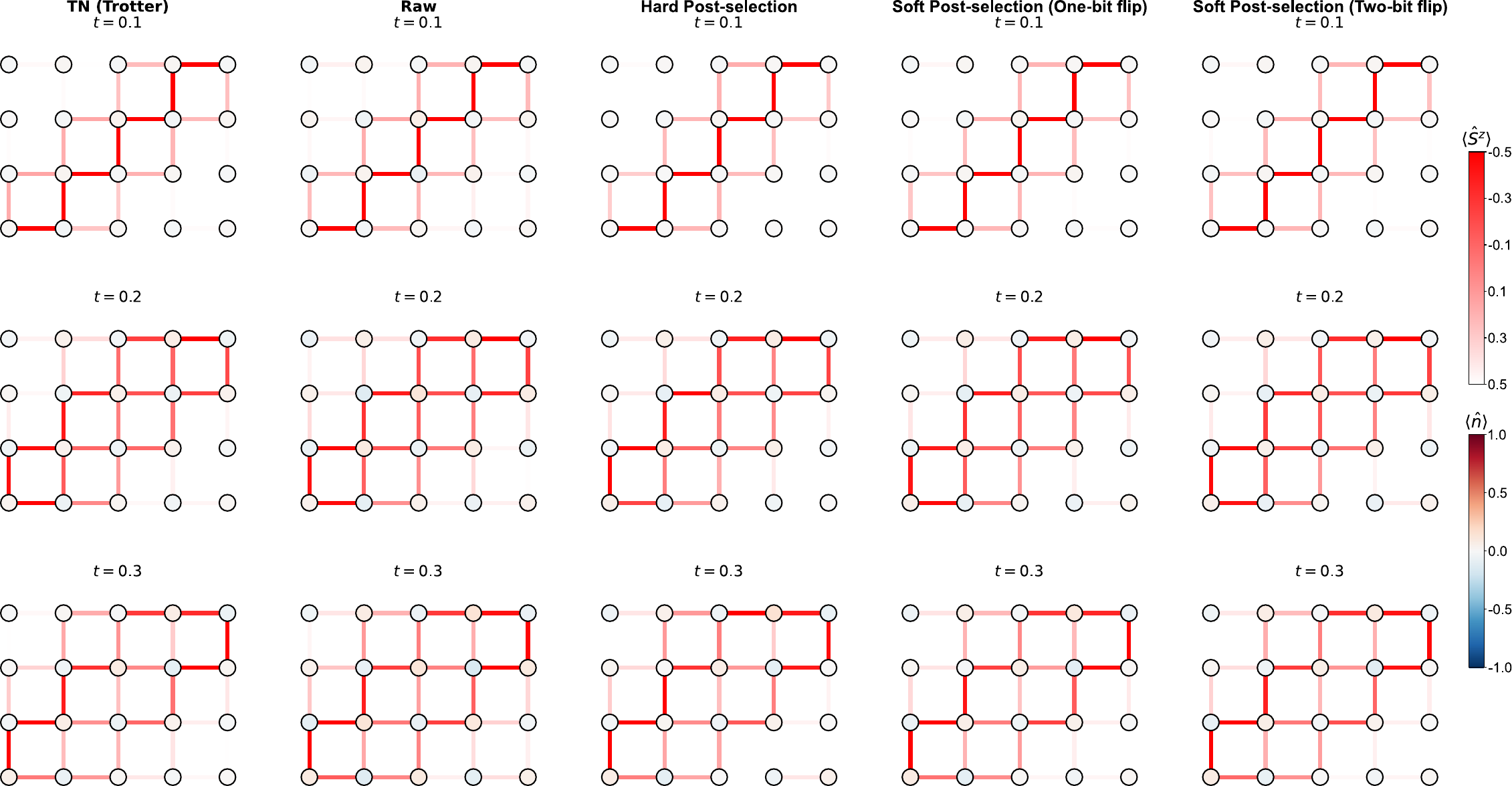}
        \caption{Snapshots of the off-resonant dynamics of an initial diagonal string on a $ 5\times4 $ lattice in the presence of a plaquette term with $J = 4$ comparing the raw hardware and different post-selected data with the Trotterized TN data.}
        \label{fig:tomography_J=4_off}
    \end{figure*}

\end{document}